\documentclass[twocolumn]{aastex631}
\usepackage{newtxtext,newtxmath}
\usepackage[T1]{fontenc}

\usepackage{graphicx,graphics}	
\usepackage{color,epsfig}
\usepackage{amsmath}	
\usepackage{amssymb}	
\usepackage{psfrag}
\usepackage[T1]{fontenc}
\usepackage{ebgaramond}
\usepackage{newtxmath}
\usepackage{booktabs} 
\usepackage{longtable}
\usepackage{supertabular}
\usepackage{textcomp}

\def\n{\hat{\bm{n}}}

\newcommand{\kk}{\boldsymbol{k}}
\newcommand{\kp}{\kk_\perp}
\newcommand{\kll}{k_\parallel} 
\newcommand{\xx}{\boldsymbol{x}}
\newcommand{\U}{\boldsymbol{U}}

\newcommand{\bfl}{\boldsymbol{\ell}}

\newcommand{\Hi}{H\,\textsc{i}}
\newcommand{\Hii}{H\,\textsc{i}\textsc{i}}
\newcommand{\HI}{H\,\textsc{i}~}

\def\a{{\bm{a}}}

\def\v{\bm{v}}

\def\V{\mathcal{V}}
\def\N{\mathcal{N}}
\def\S{{\mathcal S}}
\def\u{{\bf U}}

\defcitealias{Gill_2024_2d3vc}{Paper~I}
\defcitealias{Gill_2025_mabs}{Paper~II}

\defcitealias{Gill_2025_mwa1}{Paper~III}

\begin{document}

\title{ The  Epoch of Reionization 21 cm  Bispectrum at $z=8.2$ from MWA data II: Smooth Component Filtering}

\correspondingauthor{Sukhdeep Singh Gill}
\email{sukhdeepsingh5ab@gmail.com}

\author[0000-0003-1629-3357]{Sukhdeep Singh Gill}
\affiliation{Department of Physics, Indian Institute of Technology Kharagpur, Kharagpur 721 302, India}

\author[0000-0003-1206-8689]{Khandakar Md Asif Elahi}
\affiliation{Centre for Strings, Gravitation and Cosmology, Department of Physics, Indian Institute of Technology Madras, Chennai 600036, India}

\author[0000-0002-2350-3669]{Somnath Bharadwaj}
\affiliation{Department of Physics, Indian Institute of Technology Kharagpur, Kharagpur 721 302, India}

\author[0000-0002-7737-1470]{Shiv K. Sethi}
\affiliation{Raman Research Institute, C. V. Raman Avenue, Sadashivanagar, Bengaluru 560080, India}

\author[0000-0002-6216-2430]{Akash Kumar Patwa}
\affiliation{Raman Research Institute, C. V. Raman Avenue, Sadashivanagar, Bengaluru 560080, India}

\begin{abstract}

The 21 cm bispectrum (BS) offers a powerful probe of the Epoch of Reionization (EoR), but its observational access is severely hindered by dominant astrophysical foregrounds. Considering Murchison Widefield Array (MWA) observations at $154.2~\mathrm{MHz}$ ($z=8.2$), we mitigate the foregrounds with Smooth Component Filtering (SCF) and estimate the 21 cm BS.   We validate the pipeline using a simulated 21 cm signal and show that the input BS is recovered for modes $k_{\parallel} \ge [k_\parallel]_f=0.135~{\rm Mpc}^{-1}$. Applied to actual data, the SCF produces substantial foreground suppression, reducing the amplitude of the cylindrical BS $B(k_{1\perp},k_{2\perp},k_{3\perp},k_{1\parallel},k_{2\parallel})$ by $3$–$4$ orders of magnitude. The artifacts due to the missing frequency channels in the data are also suppressed. The resulting EoR window is significantly cleaner at small  $k_{\perp}$. We adopt the region $(k_{1 \perp},k_{2 \perp},k_{3 \perp})\leq 0.026~{\rm Mpc}^{-1}$ and $(k_{1\parallel},k_{2\parallel},k_{3\parallel})>0.135~{\rm Mpc}^{-1}$ to evaluate the 3D spherical BS and constrain the EoR signal. By combining estimates over all triangle shapes, we place the lower and upper limits on the mean cube brightness temperature fluctuations $\Delta^3$. The estimates are consistent with statistical fluctuations from system noise. The most stringent lower limit $\Delta^3_{\rm LL}=-(1.25\times 10^4)^3~{\rm mK}^3$ and upper limit $\Delta^3_{\rm UL}=(1.22\times 10^4)^3~{\rm mK}^3$ are obtained at $k_1=0.281~{\rm Mpc}^{-1}$. Additional observing time will reduce the noise level and enable substantially tighter constraints on the EoR signal.

\end{abstract}



\section{Introduction}

The Epoch of Reionization (EoR), when the hydrogen in the intergalactic medium (IGM) transitioned from the neutral (\Hi) to the ionized (\Hii) state, marks an important transition in cosmological evolution that is closely related to galaxy formation and several other related  astrophysical processes.   The 21 cm signal originating from the spin-flip transition of the \Hi\, is a promising probe of the rich EoR physics. Considerable observational efforts are underway to extract this information by measuring the power spectrum (PS) of the 21 cm brightness temperature fluctuations using radio telescopes such as the upgraded Giant Metrewave Radio Telescope (uGMRT)\footnote{\url{http://www.gmrt.ncra.tifr.res.in/}} \citep{Swarup_1991,Gupta_2017}, Murchison Widefield Array (MWA)\footnote{\url{https://www.mwatelescope.org/}} \citep{tingay13}, Low Frequency Array (LOFAR)\footnote{\url{https://www.astron.nl/telescopes/lofar/}} \citep{haarlem},  and Hydrogen Epoch of Reionization Array (HERA)\footnote{\url{https://reionization.org/}} \citep{DeBoer_2017}, and the upcoming Square Kilometre Array (SKA)-Low\footnote{\url{https://www.skao.int/}} \citep{Koopmans_2015}. 
Despite these efforts, we still do not have a detection of the EoR 21 cm PS, with the major obstacle posed by the astrophysical foregrounds that are $4-5$ orders of magnitude brighter than the expected 21 cm signal. At present, the different instruments only provide upper limits on the 21 cm PS.  The best upper limit on the mean-squared 21 cm brightness temperature fluctuations comes from the HERA experiment, which reported $[\Delta^2]\leq(21.4)^2$ mK$^2$ at $k = 0.34$ $h$ Mpc$^{ -1}$ at $z = 7.9$ \citep{HERA2023}. The high volume of data from the upcoming SKA Observatory (SKAO) is expected to yield tighter upper limits and potentially detect the signal. 

The PS can only encapsulate the full information of a Gaussian random field. Nonetheless, the 21 cm signal originating from the EoR exhibits strong non-Gaussianity, attributed to the emergence and growth of ionized regions \citep{BP2005,mondal2015, suamnm2018}. To capture the additional information encoded in the non-Gaussian features, we need to consider higher-order statistics. The bispectrum (BS) is the lowest-order statistic sensitive to non-Gaussianity and also captures mode coupling. Theoretical models and simulations have shown that the BS captures crucial information about the EoR 21 cm signal that is missed by the PS \citep{Yoshiura_2015, Shimabukuro_2016,Majumdar_2018, trott2019,Hutter_2019, Majumdar_2020, Saxena_2020, Watkinson_2021,Murmu:2023fgv, gill_eormulti, raste_2023}. It offers a powerful means to refine theoretical models and probe the underlying IGM physics. Measuring the 21 cm BS during the EoR yields tighter constraints on reionization parameters \citep{shiam2017,Watkinson_2022,tiwari_2022} and reveals the IGM heating and ionization history  \citep{Watkinson_2021,Kamran_2021b,kamran_2022,gill_eormulti}. 
As the \HI topology evolves during the EoR, the BS undergoes two characteristic sign reversals—first, early in the EoR when ionized bubbles begin to emerge in largely neutral IGM \citep{suamnm2018}, and second, near the end of reionization as the \HI\ regions fragment into isolated islands embedded in a largely ionized background \citep{raste_2023,gill_eormulti}. Moreover, the quadrupole moment of the BS is especially sensitive to the reionization scenario, and it can discriminate between “inside-out” reionization, where ionized regions grow outward from high-density sources, and “outside-in” scenarios, in which low-density regions ionize first \citep{gill_eormulti}.

The observational study of the EoR 21 cm BS is still in its infancy. An earlier work by \citet{trott2019} has analyzed 21 hours of $167$–$197$ MHz  ($z = 6.2$ – $7.5$) MWA Phase II EoR project data to estimate the 21 cm BS for a few triangle configurations and present the upper limit on the BS to be $\sim 10^{12} ~\mathrm{mK}^3~\mathrm{Mpc}^6$ on large scales. In a series of work advancing the observational BS studies, we first developed a fast estimator for the angular BS from single-frequency radio-interferometric visibility data (\citealt{Gill_2024_2d3vc}; hereafter \citetalias{Gill_2024_2d3vc}). We subsequently generalized this estimator to multifrequency observations and introduced the \texttt{AMBER} estimator, which computes the multi-frequency angular BS (MABS) and the 3D 21 cm BS from visibility data (\citealt{Gill_2025_mabs}; hereafter \citetalias{Gill_2025_mabs}).

In a recent work \cite{Gill_2025_mwa1} (hereafter \citetalias{Gill_2025_mwa1}), using the \texttt{AMBER} estimator, we have analyzed the 17 minutes of single pointing of MWA Phase II drift-scan observation centered at frequency $\nu_c = 154.2$ MHz over the redshift range $[7.4,9.2]$, with a central redshift of $z = 8.2$. That extensive work has provided the estimates for the 21 cm BS across all possible triangle configurations. We have found that the cylindrical BS exhibits a ``foreground wedge'' that is very similar to the cylindrical PS. As the foregrounds are mostly restricted to a wedge, we have shown that there is an ``EoR window'' in the cylindrical BS. Nonetheless, the periodic pattern of missing frequency channels in the MWA data is found to cause severe foreground leakage in the EoR window, and the estimates presented in \citetalias{Gill_2025_mwa1} are dominated by foregrounds. The best $2\sigma$ upper limits on mean cube brightness temperature fluctuations were found to be $\Delta^3_{\rm UL} = (1.81\times 10^3)^3~\mathrm{mK}^3$ at $k_1 = 0.008~\mathrm{Mpc}^{-1}$ and $\Delta^3_{\rm UL} = (2.04\times 10^3)^3~\mathrm{mK}^3$ at $k_1 = 0.012~\mathrm{Mpc}^{-1}$ for equilateral and squeezed triangles, respectively. Mitigating foreground contamination in the data is crucial for advancing toward the detection of the signal.


In the present work, we mitigate foreground contamination using the Smooth Component Filtering (SCF) technique. The SCF has been introduced earlier to constrain the 21 cm EoR PS by \citet{Elahi_missing}, who have demonstrated that the SCF mitigates foreground contamination, reduces leakage in the EoR window, and eliminates artifacts arising from missing channels. Here, we have used the SCF on the same observational data that was used in \citetalias{Gill_2025_mwa1}, and have utilized the \texttt{AMBER} estimator on the residual-filtered visibilities to compute the 21 cm BS. We report improved upper limits and also present the lower limits on the EoR 21 cm BS.

The paper is organized as follows. In Section \ref{sec:data}, we describe the observational data and the simulation of system noise. Section \ref{sec:revise} briefly revisits our BS estimation pipeline. Section \ref{sec:SCF} provides the details of the SCF technique, its validation, and results obtained by implementing it on the actual observational data. Section \ref{sec:results} presents the statistics of the estimated BS, the parameterization of the spherical BS, final results of the mean cube brightness temperature fluctuations, and constraints on the EoR BS signal. Finally, Section \ref{sec:sum} summarizes the findings and concludes the paper.

The cosmological parameters used in this work are taken from \citet{planck2020}.

\section{Data}
\label{sec:data}

We used data from the MWA Phase II drift-scan observations (Project ID: G0031) described in \citet{Patwa_2021}. The same dataset was used in \citet{chatterjee_2024} and \citet{Elahi_missing}  to estimate the multifrequency angular power spectrum (MAPS) and the cylindrical PS, and also used in \citetalias{Gill_2025_mwa1} to compute the MABS and the 3D BS. The data corresponds to a single pointing center at $ { \rm (RA, DEC)} =(6.1^\circ,-26.7^\circ)$, which is observed for a total of $\approx 17$ minutes across  $9$ nights and $11$ time stamps (refer to \citetalias{Gill_2025_mwa1} for details). The observation comprises $N_c = 768$ frequency channels with a frequency resolution of $\Delta\nu_c = 40$ kHz, resulting in a total bandwidth of $B_{\rm bw} = 30.72$ MHz, centered at frequency $\nu_c = 154.2$ MHz. This dataset probes the \HI 21 cm line across the redshift range $[7.4,9.2]$, with a central redshift of $z_c = 8.2$. The full bandwidth is divided into $24$ coarse bands, each containing $32$ fine channels. Within each coarse band, four edge channels at both ends and one central channel are flagged. The resulting flagging pattern is periodic with a frequency spacing of $1.28$ MHz. In addition, different channels may be flagged at each baseline to avoid radio frequency interference (RFI). The baselines in the data are mostly ($\sim99\%$) within $U = 250$ \citep{chatterjee_2024}. Here, we restrict our analysis to baselines in the range  $5 \le U \le  277$. For each baseline $\u$ and frequency $\nu$, separate measurements are available for two orthogonal polarizations, namely XX and YY, respectively,  denoted  $\V_{\rm XX}(\u, \nu)$ and $\V_{\rm YY}(\u, \nu)$. We treat these as independent measurements and combine them when gridding the visibility data.

To quantify the system noise contribution to the BS estimates, we perform dedicated ``noise-only'' simulations. We have modeled the noise in the real and imaginary parts of the visibilities as Gaussian random variables \citep{Thompson_1986, chatterjee_2024} with zero mean and standard deviation computed using,
\begin{equation}\label{eq:noise}
    \sigma_{\rm N} = \dfrac{1}{\eta_Q} \dfrac{\sqrt{2}k_BT_{\rm sys}}{A_{\rm eff}\sqrt{\Delta\nu_c t_{\rm int}}}\,,
\end{equation}
where the effective area $A_{\rm eff}=21.4\,{\rm m}^2$, system temperature $T_{\rm sys} = 400$ K, quantization efficiency $\eta_Q=1$,  and the integration time $t_{\rm int}=90$ sec, yielding $\sigma_{\rm N} \approx 20~\mathrm{Jy}$ \citep{Patwa_2021}.
Note that at low frequencies  $T_{\rm sys}$ is dominated by $T_{\rm sky}$, the sky temperature.   
We have generated noise for each baseline, frequency channel, timestamp, and polarization, and assumed it to be uncorrelated. Simulated visibilities are generated using the same baseline and frequency configurations as the actual data, and incorporate identical flagging information. To ensure statistical robustness, we produce 100 independent noise realizations. These simulations are used to quantify the uncertainty in the measured BS due to system noise.


\section{Revisiting the estimator}
\label{sec:revise}
Our analysis here employs the \texttt{AMBER}, which is a visibility-based 21 cm BS estimator that we used in our earlier analysis of the same MWA data  \citepalias{Gill_2025_mwa1}. The estimator approximates the MWA primary beam pattern $A(\theta)$ as a Gaussian, i.e. $A(\theta)=\exp(-\theta^2/\theta_0^2)$ and $\theta_0=0.6\theta_F$, where $\theta_{\rm F}\approx24.68^\circ$ is the FWHM of the MWA primary beam \citepalias{Gill_2024_2d3vc,Gill_2025_mabs}. In this section, we briefly revisit the formalism of the \texttt{AMBER}, the 21 cm BS estimator. 
 The reader is referred to \citetalias{Gill_2025_mabs} for details of the estimator.

The estimator operates through a three-step procedure, which we briefly describe below.

\textbf{I. Gridding the visibilities:} First, we grid  the radio-interferometric visibility data 
$\V(\u_i,\nu)$. For each frequency channel, we introduce a square grid with a spacing of $\Delta U_g= \sqrt{\ln2}/{\pi\theta_0}\approx 1$ in the $\u$-plane, and assign each visibility $\V(\u_i,\nu)$ to the grid point $\u_g$ nearest to $\u_i$ using
\begin{equation}
\V_{g}(\nu) = \sum^{N_g(\nu)}_{i}\tilde{w}(\u_g-\u_i) \, \V(\u_i,\nu) \, F(\u_i,\nu).
\label{eq:vg}
\end{equation}
 Here, $\tilde{w}(\u_g-\u_i)=1$ for the nearest grid point and $=0$ otherwise.  The factor  $F(\u_i, \nu)$  accounts for any flagged data, taking a value of \( 0 \) if the data at baseline $\u_i$  and frequency $\nu$ is flagged, and $1$ otherwise. We also define  $N_g(\nu)=\sum\tilde{w}(\u_g-\u_i)  \, F(\u_i,\nu) $, which quantifies the number of visibilities contributing to any grid point $g$ at frequency $\nu$.

\textbf{II. MABS estimation:} In the second step, we compute the MABS $B_A({\ell_1}, {\ell_2}, {\ell_3},\nu_1,\nu_2,\nu_3)$ by utilizing its relation with three visibility correlations, given as 
\begin{equation}
\begin{aligned}
&& \langle \V(\u_1,\nu_1) \V(\u_2,\nu_2) \V(\u_3+\Delta \u,\nu_3)\rangle =
 (\pi\theta_0^2 Q^3/3)  \\
&&\times
  \exp\left [-{\pi^2\theta_0^2\Delta U^2/3} \right ] 
 B_A({\ell_1}, {\ell_2}, {\ell_3},\nu_1,\nu_2,\nu_3) \,.
\label{eq:mabs}
\end{aligned}
\end{equation}
where the Rayleigh-Jeans factor, $Q=2 k_B/\lambda^2$, converts brightness temperature to specific intensity. The values of $\theta_0$ and $Q$ are fixed at the central frequency $\nu_c$, assuming the bandwidth is sufficiently small.
The three baselines $(\u_1, \u_2, \u_3)$ form a closed triangle i.e., $\u_1 + \u_2 + \u_3=0$, where $\bfl=2\pi \u$ and $\Delta \u$ is the deviation from a closed triangle configuration.

The estimator, \texttt{AMBER}, evaluate Eq.~(\ref{eq:mabs}) on the gridded visibilities $\V_{g}(\nu)$. For each frequency channel, the gridded $\u$ plane is divided into several concentric annular rings. An estimate of the binned MABS $B_A(\ell_1, \ell_2, \ell_3, \nu_1, \nu_2,\nu_3)$ is obtained from the combination of three annular rings of radius $(\ell_1, \ell_2, \ell_3)$  respectively at the frequencies ($\nu_1, \nu_2,\nu_3$). Furthermore, we assume that the 21 cm signal is ergodic along the frequency axis, or equivalently, along the line-of-sight (LoS) direction. As a result, the MABS depends only on the frequency separations, i.e., we can then express the MABS as $B_A(\ell_1, \ell_2, \ell_3, \Delta\nu_1, \Delta\nu_2)$, where $\Delta\nu_1 = \nu_1 - \nu_3$ and $\Delta\nu_2 = \nu_2 - \nu_3$.

\textbf{III. 3D BS estimation:} In the third step, we obtain the 3D cylindrical BS $B(k_{1 \perp},k_{2 \perp},k_{3 \perp},k_{1\parallel},k_{2\parallel})$ by performing a 2D Fourier transform of the MABS over $(\Delta\nu_1, \Delta\nu_2)$, as given by \citet{Bharadwaj2005},
\begin{equation}
    \begin{aligned}
 &B(k_{1 \perp},k_{2 \perp},k_{3 \perp},k_{1\parallel},k_{2\parallel}) = 
 r^4 r'^2 \int d(\Delta \nu_1) \int d(\Delta \nu_2)  
\\&\times \exp{[-ir'(k_{1 \parallel}\Delta \nu_1+k_{2 \parallel}\Delta \nu_2)]} \, B_A(\ell_1, \ell_2, \ell_3, \Delta \nu_1, \Delta \nu_2)~,
\end{aligned}
\label{eq:3dbs_mabs}
\end{equation}
where $r$ is the comoving distance corresponding to the 21 cm signal received at $\nu_c$, which can be calculated from the redshift $z_c=\left(\frac{1420 ~\rm{MHz}}{\nu_c}\right)-1$, and $r^\prime=\frac{dr}{d\nu}$. For the present data at $z_c=8.2$, we have $r\simeq9210 {~\rm Mpc}$ and $r^\prime\simeq17~{\rm Mpc~MHz^{-1}}$. The cylindrical BS $B(k_{1 \perp},k_{2 \perp},k_{3 \perp},k_{1\parallel},k_{2\parallel})$ is a function of five independent variables, where $\kp={\bfl}/{r}$ and $\kll$  are respectively the components of $\kk$  perpendicular and parallel to the LoS direction. The $k_{3\parallel}$ component can be determined by the closed triangle condition $k_{3\parallel}=-k_{1\parallel}-k_{2\parallel}$. The 3D cylindrical BS exhibits ripples along  $(k_{1\parallel},k_{2\parallel})$ that originate from the Fourier transform in Eq.~(\ref{eq:3dbs_mabs}), caused by the abrupt truncation of the MABS at the frequency domain boundaries.  We introduce a 2D Blackman–Nuttall window function \citep{Nuttall_1981} that smoothly tapers the MABS at the boundaries, which suppresses the edge effects and mitigates the ripples along $(k_{1\parallel},k_{2\parallel})$. 

\citetalias{Gill_2025_mwa1} has implemented the \texttt{AMBER}---namely steps \textbf{I, II} and \textbf{III}---to analyze the same dataset considered in the present paper and described here in Section~\ref{sec:data}. \citetalias{Gill_2025_mwa1} presents measurements of the 3D BS for all possible triangle configurations available from the data, along with a detailed investigation of the ``foreground wedge'' and the ``EoR window'' for the 21 cm BS. It was found that, similar to the estimated 21 cm PS  \citep{chatterjee_2024}, the estimated  21 cm BS exhibits substantial foreground leakage into the EoR window. A detailed analysis of the 21 cm PS \citep{Elahi_missing} has shown that this leakage arises due to a combination of the strong frequency-dependent foreground contamination present in the data, and the periodic pattern of missing frequency channels characteristic of the MWA, and they have proposed Smooth Component Filtering (SCF) as a technique to mitigate this contamination. 

In the present work, we have applied the SCF to the gridded data. This appears as an additional step between steps \textbf{I} and \textbf{II} discussed above.

\section{Smooth Component Filtering}
\label{sec:SCF}

The details of the SCF technique can be found in \cite{Elahi_missing}; here, we provide a brief overview. After gridding the visibility data $\V_{g}(\nu)$ in step I in Section \ref{sec:revise}, we calculate its smooth component through a convolution
\begin{equation}
    \V^s_{g}(\nu_n) = (  \V_{g} * H )(\nu_n) = \sum_{m=1}^{N_c} \V_{g}(\nu_{m}) H(n-m)\,,
    \label{eq:conv}
\end{equation}
where $\V_{g}(\nu)$ is convolved with a Hann window function $H(n)$, defined as 
\begin{equation}
    H(n) = \frac{1}{4N} \left[ 1 + \cos \left( 2\pi \frac{n}{2N} \right)   \right] \, , \, \,  -N \leq n \leq N \,,
    \label{eq:Hann}
\end{equation}
and $H(n)=0$ otherwise.  The smoothing scale in $H(n)$ is set by the width of the window $2 N +1$. Here we have used $N=50$, which corresponds to a smoothing scale of half-width $2\, {\rm MHz}$. The rationale for the choice of the $2\, {\rm MHz}$ has been discussed in detail in \cite{Elahi_missing}.  Next, we obtain the filtered component by
\begin{equation}
    \V^f_{g}(\nu) =\V_{g}(\nu) - \V^s_{g}(\nu) \,.
    \label{eq:res}
\end{equation}
and use this in step II of Section \ref{sec:revise} to estimate the MABS (Eq.~\ref{eq:mabs}).  We note that the missing channels also affect the convolution (Eq.~\ref{eq:conv}), and to ensure a correct normalization, we have divided $\V^s_{g}(\nu)$ obtained from Eq.~(\ref{eq:conv}) with $(F * H)(\nu)$,  where  $F$ is the flagging function defined earlier. Also, the convolution fails at the edges of the band, and we have discarded  $N$ channels from both ends of $\V^f_{g}(\nu)$ to account for this. The final MABS is obtained over the bandwidth $B^f_{\rm bw} = B_{\rm bw}-2N\Delta\nu_c=26.72~{\rm MHz}$. 

In summary, we first grid the visibilities (Eq.~\ref{eq:vg}), then obtain the filtered visibility data (Eq.~\ref{eq:res}), and use it to compute the MABS (Eq.~\ref{eq:mabs}) and the 3D BS (Eq.~\ref{eq:3dbs_mabs}).  Before implementing the estimator on actual data, we first validate it using a simulated 21 cm signal to ensure that it does not cause any signal loss beyond the smoothing scale.

\subsection{Validation}

We simulate visibility data for a non-Gaussian sky signal. First, we generate a Gaussian random field $\delta T_{\rm G} (\xx)$ using the input model PS $P(k)=k^{-2}$, and then obtain the non-Gaussian field $\delta T_{\rm b} (\xx)$ using a local non-linear transformation, $\delta T_{\rm b} (\xx)= \delta T_{\rm G} (\xx) + \,\left[\delta T^2_{G} (\xx)- \langle\delta T^2_{G} (\xx)\rangle \right] \,$. We use $\delta T_{\rm b} (\xx)$ to simulate the visibilities on the same MWA baseline distribution as the actual data (implementing Eq.~4 of \citetalias{Gill_2024_2d3vc}), and generated $100$ such realizations. The reader is referred to \citetalias{Gill_2025_mabs} for details of the simulations. 



\begin{figure}[htbp]
  \centering
  \includegraphics[width=.5\textwidth]{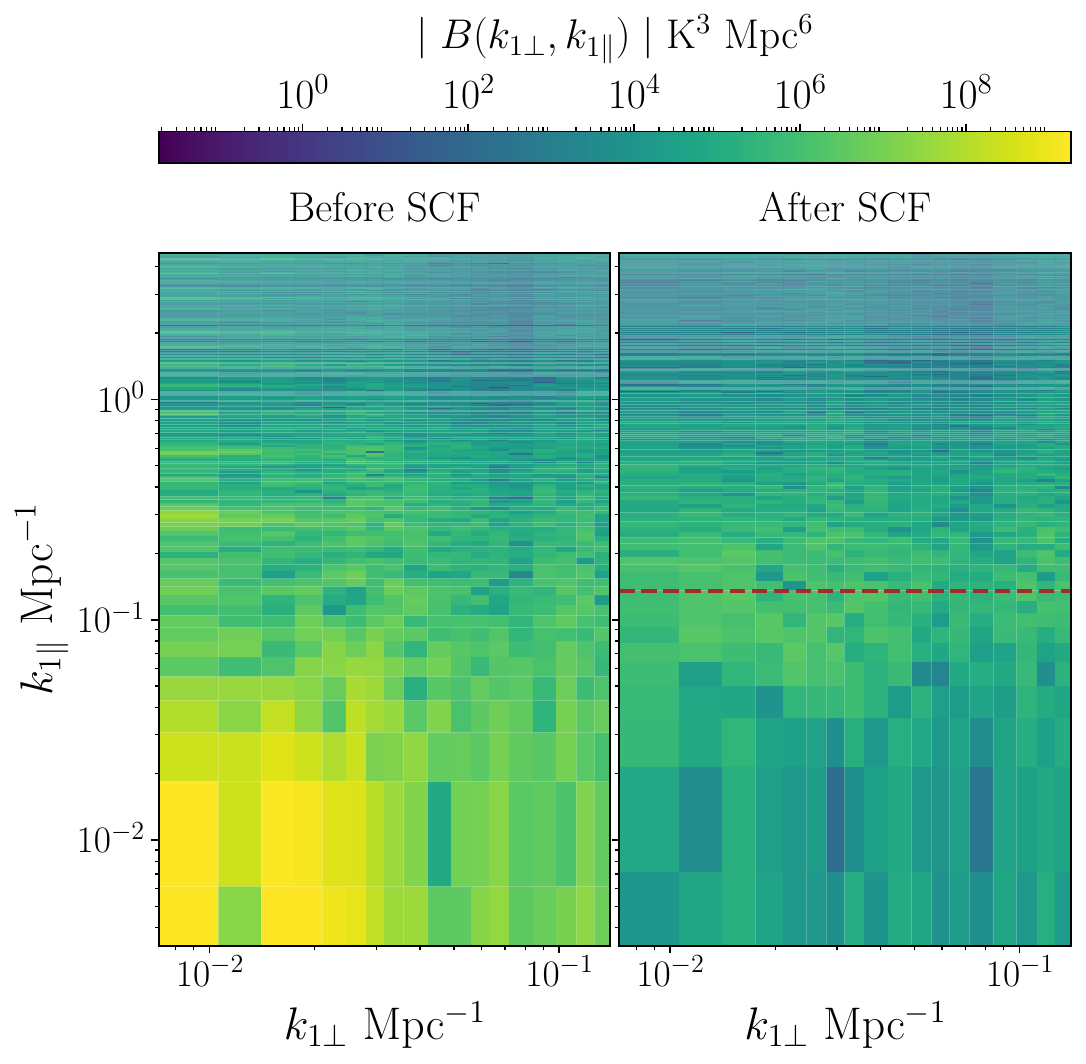}
  \vspace{4mm} 
  \includegraphics[width=.5\textwidth]{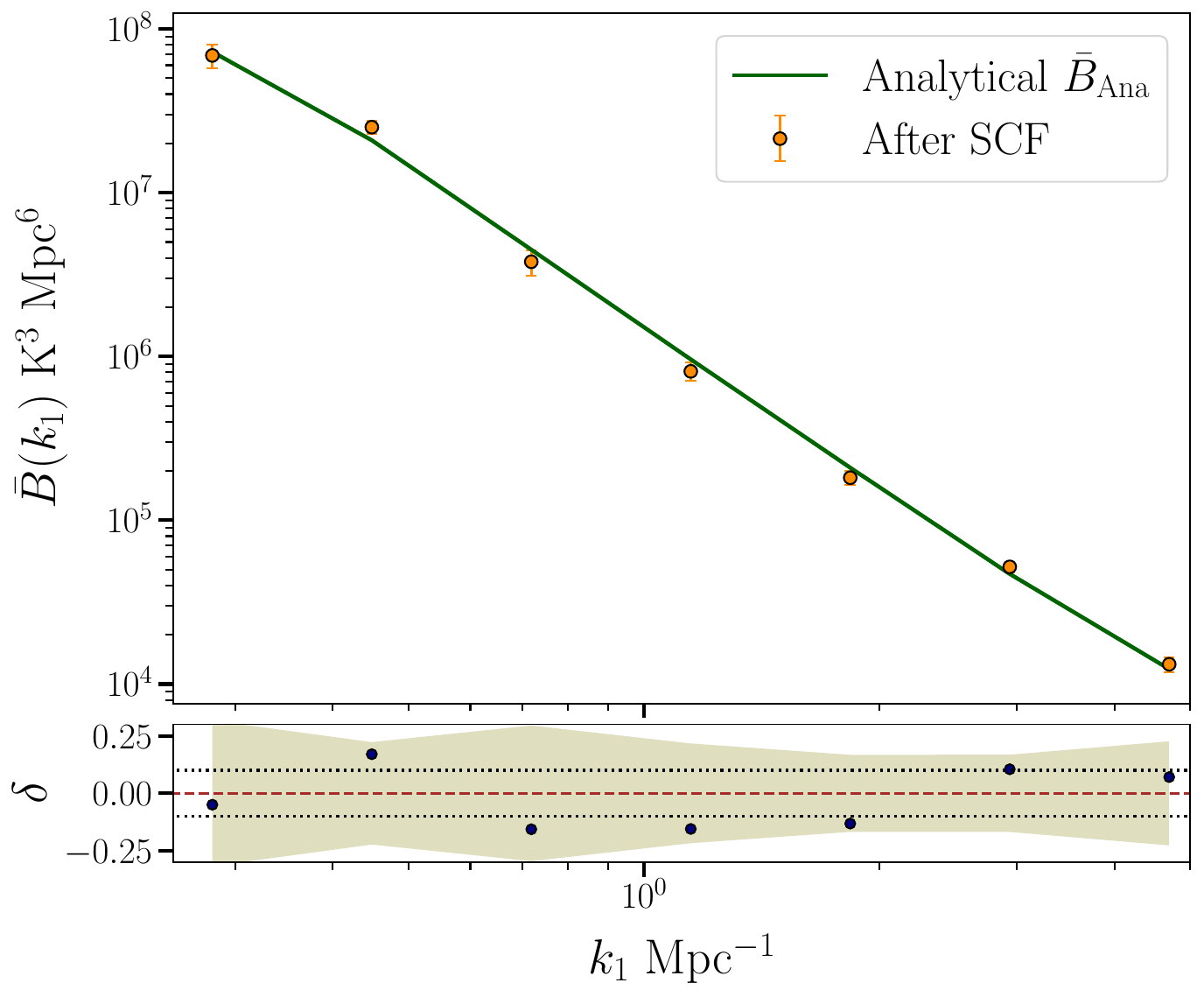}
  \caption{Validating the SCF. The top panels show the cylindrical BS $B(k_{1\perp},k_{2\perp},k_{3\perp},k_{1\parallel},k_{2\parallel})$ plotted as a function of $(k_{1\perp},k_{1\parallel})$ for the cases before (left) and after (right) applying the SCF. Here we consider the equilateral triangles, $k_{1\perp}\simeq k_{2\perp}\simeq k_{3\perp}$, and set $k_{1\parallel}=k_{2\parallel}=-k_{3\parallel}/2$. The horizontal red dashed line marks the scale $[k_{\parallel}]_f=0.135\ \mathrm{Mpc}^{-1}$; the signal in the $k_{1\parallel}$   modes below this line is effectively suppressed by the filter. The middle panel shows the 3D spherical BS $\bar{B}$ as a function of $k_1$, obtained by combining estimates for all triangle shapes. The orange markers indicate the values measured from the simulations after the SCF, while the green solid line shows the analytical prediction for the BS $\bar{B}_{\rm Ana}$ of the simulated signal. The error bars on the estimates are obtained from $100$ independent realizations of the simulations. The bottom panel shows the fractional deviation, $\delta=(\bar{B}-\bar{B}_{\rm Ana})/\bar{B}_{\rm Ana}$. The brown-dashed line corresponds to the zero level, while the two black-dotted lines represent $\delta=\pm0.1$. The shaded region shows expected $2\sigma$ statistical fluctuations. 
 }
  \label{fig:validation}
\end{figure}

Fig.~\ref{fig:validation} shows the results for the 3D BS obtained from the simulated visibility data, both before and after applying the SCF. The top row presents the cylindrical BS $B(k_{1 \perp},k_{2 \perp},k_{3 \perp},k_{1\parallel},k_{2\parallel})$ as a function of $(k_{1 \perp},k_{1\parallel})$, considering equilateral triangles $k_{1\perp}\approx k_{2 \perp}\approx k_{3\perp}$ and $k_{1\parallel}=k_{2\parallel}=-k_{3\parallel}/2$. The left and right panels show the results before and after the SCF, respectively. 
Although the SCF effectively filters out the smooth frequency-dependent component from $\V_{g}(\nu)$, it suppresses the 21 cm signal at small $k_{\parallel}$; however, we expect to recover the 21 cm signal at large $k_{\parallel}$ where $k_{\parallel} \ge [k_{\parallel}]_f$. \citet{Elahi_missing} have used a conservative value $[k_{\parallel}]_f=2 \pi/(r^{'} \, 2 \, {\rm MHz})=0.185 \, {\rm Mpc}^{-1}$, which corresponds to the half-width of the Hann filter. However, it may be argued that  $[k_{\parallel}]_f=0.093 \, {\rm Mpc}^{-1}$, corresponding to the full-width of the Hann filter, is a more appropriate choice, as it allows the 21 cm signal to be probed over a broader $k_{\parallel}$ range.  Simulations show that the range $k_{\parallel}\ge [k_{\parallel}]_f=0.135~{\rm Mpc}^{-1}$ can be used to reliably estimate the 21 cm PS without significant signal loss due to the  SCF \citep{sarkar_2026}. 
For the present analysis, we adopted $ [k_{\parallel}]_f=0.135~{\rm Mpc}^{-1}$, indicated by the red dashed line in the upper right panel of  Fig.~\ref{fig:validation}. The comparison of the cylindrical BS before and after applying the SCF shows that the values of $B(k_{1\perp}, k_{1\parallel})$ appear to remain unchanged in the region  $k_{\parallel}\ge   [k_{\parallel}]_f$, which we use to estimate the 21 cm BS. In contrast, we see that the SCF largely erases the signal below the dashed line. 
Furthermore, \citet{Elahi_missing} have noted, in the context of the PS, that the SCF is not effective at the large baselines, where the oscillations due to baseline migration have a period shorter than the $2 \, {\rm MHz}$  smoothing scale employed by the SCF. 
 Consistent with their findings, the BS analysis of the actual data in later sections also shows relatively strong foreground contamination at larger $k_{1 \perp}$, even after applying the SCF. Motivated by this, we only utilize the modes satisfying $(k_{1 \perp},k_{2 \perp},k_{3 \perp})\leq 0.026~{\rm Mpc}^{-1}$, together with the condition $(k_{1\parallel},k_{2\parallel},k_{3\parallel})>[k_{\parallel}]_f$, to estimate the 21 cm BS.

Further, we spherically bin the BS $B(k_{1 \perp},k_{2 \perp},k_{3 \perp},k_{1\parallel},k_{2\parallel})$ estimates to evaluate the spherical BS $\bar{B}(k_1,k_2,k_3)$, where $k_1=|\kk_1|=\sqrt{k_{1\perp}^2 + k_{1\parallel}^2}$, etc., and $k_1 \ge k_2 \ge k_3$. 
  The middle panel of Fig.~\ref{fig:validation} shows $\bar{B}(k_1,k_2,k_3)$ as a function of $k_1$, the triangle size. We have combined the estimates corresponding to all $(k_2,k_3)$ values, essentially adding results across all triangle shapes.  The orange markers are the estimates obtained from the simulated visibilities after the SCF, and the error bars are computed from 100 realizations of the simulations. The solid green line represents the analytical predictions of the BS $\bar{B}_{\rm Ana}$ obtained using Eq.~(21) of \citetalias{Gill_2025_mabs}. Note that the analytical predictions do not include any information on the baseline distribution, missing frequency channels, or the SCF. We see that the input signal is recovered with very high accuracy. The bottom panel shows the fractional deviation of the estimates from the analytical predictions, $\delta=(\bar{B}-\bar{B}_{\rm Ana})/\bar{B}_{\rm Ana}$. The fractional deviations are below $20\%$ over the entire range of $k_1$. The shaded region shows the expected $2\sigma$ statistical fluctuations. We see that the fractional deviations are well within the predicted $2 \sigma$ statistical fluctuations,  validating our estimator.

\subsection{Implementation on Data}
We implement the BS estimation procedure on the actual observational data and discuss the results at each step below.

\subsubsection{MABS}
\begin{figure*}
\centering
\includegraphics[width=1\textwidth]{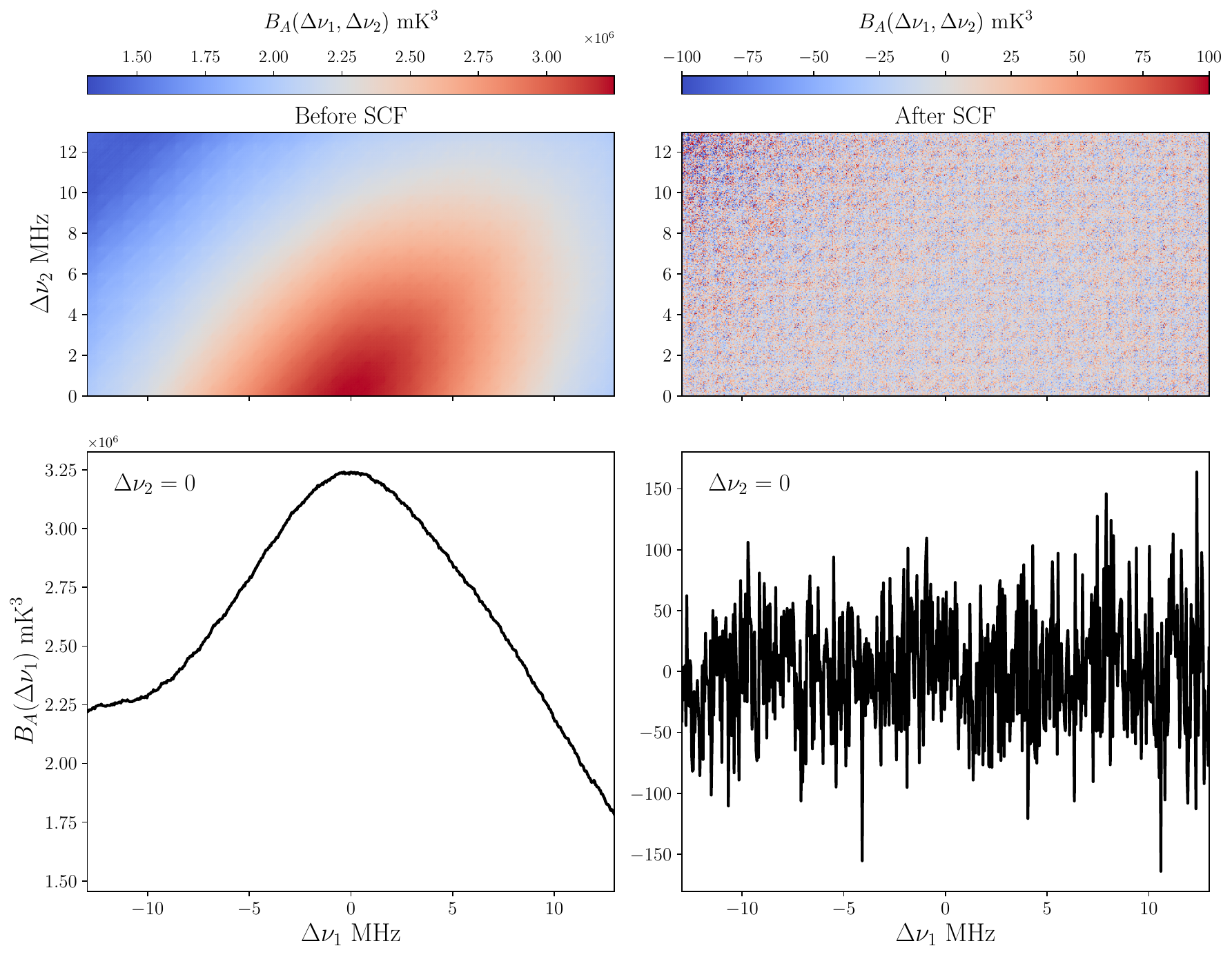}
\caption{ MABS
$B_A(\ell_1,\ell_2,\ell_3,\Delta\nu_1,\Delta\nu_2)$ before SCF (left column) and after SCF (right column)
for the equilateral configuration $\ell_1=\ell_2=\ell_3=210$. 
Top panels: $B_A$ plotted as a function of $(\Delta\nu_1,\Delta\nu_2)$. 
Bottom panels: one-dimensional slices of the corresponding top-row heatmaps, showing $B_A$ as a function of $\Delta\nu_1$ at fixed $\Delta\nu_2=0$.
 }
\label{fig:mabs}
\end{figure*}

For the current dataset, we estimate the MABS, $B_A(\ell_1,\ell_2,\ell_3,\Delta\nu_1,\Delta\nu_2)$ for 647 distinct $(\ell_1,\ell_2,\ell_3)$ triplets that together cover all triangle configurations possible with this data. For each triangle $(\ell_1,\ell_2,\ell_3)$, we have the estimates on a regular grid in the $(\Delta\nu_1,\Delta\nu_2)$ plane with spacing $\Delta\nu_c = 40\ \mathrm{kHz}$ over the range $-B^f_{\rm bw}$ to $B^f_{\rm bw}$. The region with $|\Delta\nu_2-\Delta\nu_1|>B^f_{\rm bw}$ remains unsampled (see Figure~2 of \citetalias{Gill_2025_mabs}), so we restrict the analysis to $-B^f_{\rm bw}/2 \le \Delta\nu_1,\Delta\nu_2 \le B^f_{\rm bw}/2$ for ease of subsequent analysis. Additionally, the MABS exhibits the symmetry $B_A(\ell_1,\ell_2,\ell_3,-\Delta\nu_1,-\Delta\nu_2)=B_A(\ell_1,\ell_2,\ell_3,\Delta\nu_1,\Delta\nu_2)$; consequently, it is sufficient to consider only the upper half-plane  $-B^f_{\rm bw}/2 \le \Delta\nu_1 \le B^f_{\rm bw}/2$ and $0 \le \Delta\nu_2 \le B^f_{\rm bw}/2$.


The top panels of Fig.~\ref{fig:mabs} show $B_A(\ell_1, \ell_2, \ell_3, \Delta\nu_1, \Delta\nu_2)$ as a function of $(\Delta\nu_1, \Delta\nu_2)$ for an equilateral configuration $\ell_1=\ell_2=\ell_3=210$. The left panel shows the results before SCF (see \citetalias{Gill_2025_mwa1} for detailed discussion), while the right panel shows the results after SCF. The SCF effectively removes the dominant foreground contribution, and the MABS amplitude falls from $\sim10^6~{\rm mK}^3$ to $\sim10^2~{\rm mK}^3$. The periodic square grid pattern seen in the right panel is a manifestation of the irregular sampling of the $(\Delta \nu_1,\Delta \nu_2)$ plane (Figure 2 of \citetalias{Gill_2025_mabs}) due to the periodic pattern of missing frequency channels present in the MWA data.  Such a pattern is also present in the left panel, but it is not visible due to the dominant, smoothly varying foreground component. 
Note that there are no missing $(\Delta \nu_1,\Delta \nu_2)$ values due to the flagging. 
The bottom panels show $B_A(\ell_1, \ell_2, \ell_3, \Delta\nu_1, \Delta\nu_2)$ as a function of $\Delta\nu_1$ at fixed $\Delta\nu_{2}=0$, which corresponds to a 1D slice along the $\Delta\nu_1$ axis of the  MABS shown in the top-row. We see that the SCF effectively removes the spectrally smooth trend in the MABS, and the residual MABS exhibits fluctuations of order $\pm10^{2}\,{\rm mK}^3$.

\subsubsection{Cylindrical BS}

\begin{figure*}
\centering
\includegraphics[width=1\textwidth]{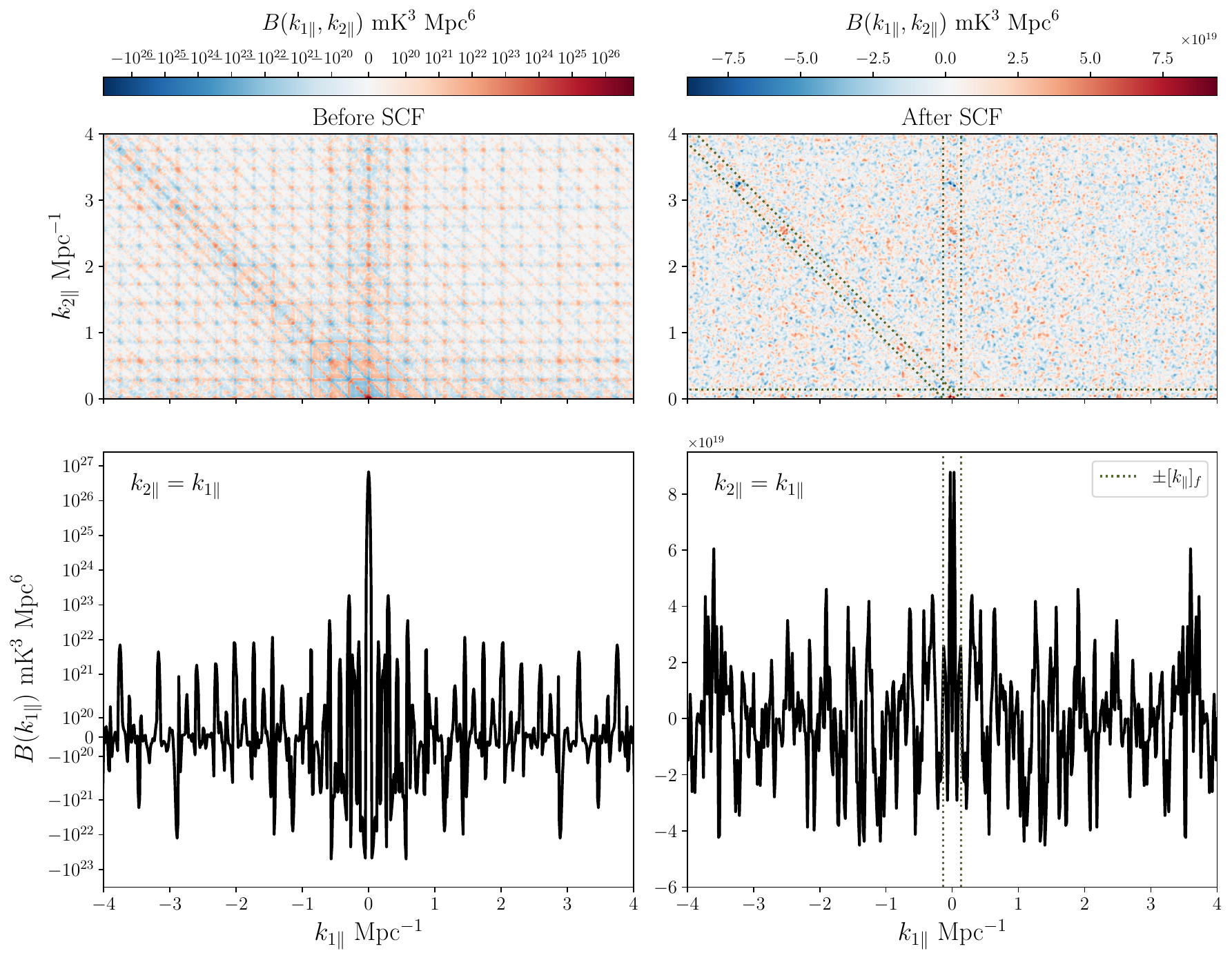}
\caption{ 
The 3D cylindrical BS
$B(k_{1 \perp},k_{2 \perp},k_{3 \perp},k_{1\parallel},k_{2\parallel})$ before SCF (left column) and after SCF (right column) for the equilateral configuration $k_{1\perp} =k_{2\perp}=k_{3\perp}= 0.023~{\rm Mpc}^{-1}$.  This is computed via a 2D Fourier transform (Eq.~\ref{eq:3dbs_mabs}) of the MABS shown in Figure~\ref{fig:mabs}.
Top panels: the BS plotted as a function of $(k_{1\parallel},k_{2\parallel})$. 
Bottom panels: one-dimensional slices of the corresponding top-row heatmaps, showing the BS as a function of $k_{1\parallel}$ with $k_{2\parallel}=k_{1\parallel}$ and $k_{3\parallel}=-2k_{1\parallel}$. The green dashed lines in the bottom-right panel mark the scales $\pm[k_{\parallel}]_f$; SCF filters out the signal from the $k_{\parallel}$ modes inside this band  (see the top panels of Figure~\ref{fig:validation}). 
}
\label{fig:cylbs}
\end{figure*}

Fig.~\ref{fig:cylbs} shows the 3D cylindrical BS $B(k_{1 \perp},k_{2 \perp},k_{3 \perp},k_{1\parallel},k_{2\parallel})$ computed from the MABS shown in Fig.~\ref{fig:mabs}. The top row shows $B(k_{1 \perp},k_{2 \perp},k_{3 \perp},k_{1\parallel},k_{2\parallel})$ as a function of $(k_{1\parallel},k_{2\parallel})$ with $(k_{1 \perp},k_{2 \perp},k_{3 \perp})=r^{-1} \, (\ell_1, \ell_2, \ell_3)=0.023~{\rm Mpc}^{-1}$ fixed. The left panel shows the results before the SCF. While a full discussion of these results can be found in \citetalias{Gill_2025_mwa1}, we provide a brief summary here. The cylindrical BS peaks at $k_{1\parallel}=k_{2\parallel}=0$ and drops rapidly at larger values of $(k_{1\parallel},k_{2\parallel})$. The bottom-left panel, which plots $B(k_{1 \perp},k_{2 \perp},k_{3 \perp},k_{1\parallel},k_{2\parallel})$ as a function of $k_{1\parallel}$ along the diagonal $k_{2\parallel}=k_{1\parallel}$, shows a peak of $\sim 10^{27}~{\rm mK^3 ~Mpc^6}$ at $k_{1\parallel}=0$. At higher $k_{1\parallel}$, the amplitude drops to $\sim 10^{21}-10^{22}~{\rm mK^3 ~Mpc^6}$. In addition to the central peak, the top left panel shows three pronounced bands aligned with $k_{1\parallel} = 0$, $k_{2\parallel} = 0$, and  $k_{1\parallel}=-k_{2\parallel}$  ($k_{3\parallel} = 0$), respectively.
Both the central peak and these dark bands are characteristic signatures of the spectrally smooth foreground contamination. Furthermore, a periodic square grid pattern extends across the entire $(k_{1\parallel}, k_{2\parallel})$ plane, arising from the periodic flagging present in the MWA data, combined with the strong foreground contamination.

Considering the results after the SCF shown in the right panels, we observe that the overall amplitude of the BS is significantly reduced. The values now fluctuate at the level of $\pm 10^{19}~{\rm mK^3 ~Mpc^6}$, as shown in the bottom-right panel. The central peak remains roughly twice the amplitude of the surrounding fluctuations at larger $k_{1\parallel}$. Nonetheless, this feature lies within the region $\pm[k_{\parallel}]_f$ (green-dotted lines), where the SCF removes the EoR 21 cm signal, and it is excluded from our final 21 cm BS estimation. Moreover, the three pronounced bands visible in the top-left panel (before SCF) are considerably diminished after the SCF. Three thin strips still persists along $k_{1\parallel} = 0$, $k_{2\parallel} = 0$, and  $k_{1\parallel}=-k_{2\parallel}$, which are likewise confined to the discarded region $\pm[k_{\parallel}]_f$. Furthermore, the periodic grid-like pattern has nearly disappeared. Together, these results demonstrate the effectiveness of the SCF in mitigating foreground contamination.

\subsubsection{Foreground Wedge}

\begin{figure*}
\centering
\includegraphics[width=1\textwidth]{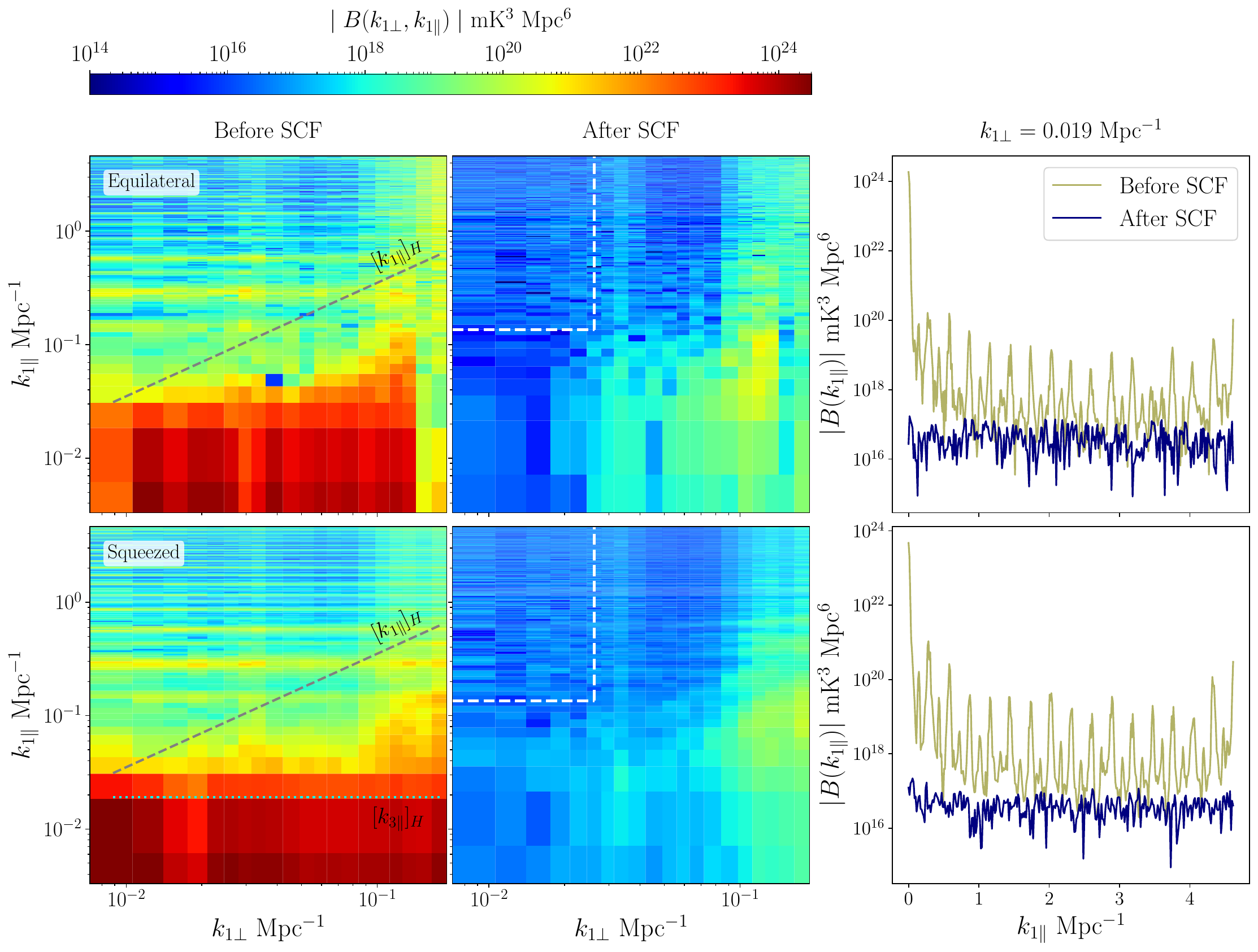}
\caption{The magnitude of the 3D cylindrical BS $\left|B(k_{1\perp},k_{2\perp},k_{3\perp},k_{1\parallel},k_{2\parallel})\right|$,  with $k_{1\parallel}=k_{2\parallel}$. The left column shows the BS as a function of $(k_{1\perp},k_{1\parallel})$ before SCF, while the middle column shows the same quantity after SCF. The top row considers equilateral configuration with $k_{1\perp}\approx k_{2\perp}\approx k_{3\perp}$ and the bottom row considers the squeezed configurations with $k_{1\perp}\approx k_{2\perp}$, $k_{3\perp}\to 0$. The dashed line in the left panels show the predicted foreground wedge boundary   $[k_{1\parallel}]_H  = [r/(r'\Delta\nu_c)]k_{1\perp}$, whereas the dotted line in the bottom left panel shows $[k_{3\parallel}]_H$   corresponding to the minimum value of $ k_{3\perp}$  $(=0.005 \, {\rm Mpc}^{-1})$. The white dashed box in the middle panels demarcates the region $k_{1 \perp}\leq 0.026~{\rm Mpc}^{-1}$ and $k_{1\parallel}>0.135~{\rm Mpc}^{-1}$, highlighting the modes we have used to evaluate the spherical BS and constrain the EoR 21 cm signal. While only two representative triangle configurations are presented here for illustration, the subsequent analysis and resulting constraints on the EoR 21 cm signal employ the full set of possible triangle configurations. The right panels show 1D sections of the heatmaps, and plot the BS as a function of $k_{1\parallel}$ at fixed $k_{1\perp} \approx 0.019~\mathrm{Mpc}^{-1}$. The green and blue lines, respectively, show results before and after SCF. }
\label{fig:wedge}
\end{figure*}

Fig.~\ref{fig:wedge} shows a comparison of the cylindrical BS
$B(k_{1\perp}, k_{2\perp}, k_{3\perp}, k_{1\parallel}, k_{2\parallel})$
before and after the SCF.  The top  panel considers equilateral triangles where $k_{1\perp}= k_{2\perp}= k_{3\perp}$ while the bottom row considers squeezed triangles where  $k_{1\perp}\approx k_{2\perp}$, and $k_{3\perp}\rightarrow 0$. In all the cases, we consider 
 $k_{2\parallel} = k_{1\parallel}$ with  $k_{3\parallel} = -2k_{1\parallel}$, which allows us to study $B(k_{1\perp}, k_{1\parallel})$. The panels in the left column show $B(k_{1\perp}, k_{1\parallel})$ before the SCF, for which the results for the equilateral triangles have been considered earlier in Fig.~5 of \citetalias{Gill_2025_mwa1}. Here, for both the equilateral and squeezed triangles,  we clearly see a foreground wedge. The dashed line shows the theoretically predicted foreground wedge boundary computed using the horizon condition $[k_{1\parallel}]_H =[{r}/{(r'\Delta\nu_c)}]\,k_{1 \perp}$.  Above this boundary $(k_{1\parallel}>[k_{1\parallel}]_H)$,  all three sides of the triangle are outside the foreground wedge, and the foreground contamination is less severe here. The interpretation of the region  $k_{1\parallel} \leq [k_{1\parallel}]_H$ is a little more complicated, as it depends on the shape of the triangle. For equilateral triangles, all three triangle sides are within the foreground wedge in this region. However, for the squeezed triangles, we also need to consider  
the dotted line $[k_{3\parallel}]_H =[{r}/{(r'\Delta\nu_c)}]\,k_{3 \perp}$ where 
$k_{3 \perp}$ corresponds to the smallest value of $k_{\perp}$ available in our data. 
The region 
$[k_{1\parallel}]_H \ge k_{1\parallel} > [k_{3\parallel}]_H$ corresponds to a situation where two sides $(\kk_1,\kk_2)$ are within the foreground wedge while the smaller side $\kk_3$ is outside the foreground wedge, whereas all three sides are within the foreground wedge for $k_{1\parallel} \le [k_{3\parallel}]_H$. For squeezed triangles, the distinction between these two regions within the foreground wedge is clearly visible in the difference in the level of foreground contamination.

We now focus on the results after the SCF,  shown in the panels of the middle column. We see that, for both triangle shapes, the foregrounds are significantly suppressed after the SCF. The suppression is particularly significant at small $k_{1\perp}$; however, it is not so effective at large  $k_{1\perp}$. This behavior after the SCF is very similar to that of the 21 cm PS presented in Fig.~4 of \citet{Elahi_missing}, to which the reader is referred for a detailed discussion. Similar considerations also hold for the 21 cm BS, and consequently,  we restrict our attention to the region $k_{1 \perp}\leq 0.026~{\rm Mpc}^{-1}$ for constraining the EoR signal. We demarcate the selected region with a white dashed lines and use the modes within this box for the subsequent analysis. Note that the results here are demonstrated for only two representative triangle configurations; nonetheless, the subsequent analysis and resulting constraints on the EoR signal employ the same selected region for the full set of possible triangle configurations. The right panels shows the 1D slice of the heatmaps along $k_{1\perp}=0.019~{\rm Mpc}^{-1}$, plotting the cylindrical BS as a function of $k_{1\parallel}$. It clearly shows a reduction in the amplitude of the BS after the SCF (blue lines) compared to the amplitude before the SCF (green line). Not only is there an overall reduction in the amplitude, but the SCF also mitigates the periodic pattern of spikes that arise due to the missing frequency channels \citet{Elahi_missing}.

\subsection{Statistical fluctuations from system noise}

We used noise-only simulations to calculate $\delta {B}_N$, the r.m.s.  statistical fluctuations expected in the estimated ${B}(k_{1\perp}, k_{2\perp}, k_{3\perp}, k_{1\parallel}, k_{2\parallel})$. The noise-only simulations are analyzed in exactly the same way as the actual data. First, we computed the MABS from each noise realization (Eq.~\ref{eq:mabs}) and then evaluated the cylindrical BS (Eq.~\ref{eq:3dbs_mabs}). From $100$ independent noise realizations, we calculated the r.m.s. of the resulting BS to obtain $\delta {B}_N$, which quantifies the expected statistical fluctuations due to system noise. In the following section, we analyze the statistics of the estimates and present the final results for the 3D spherical BS.

\section{Results}
\label{sec:results}

\subsection{Spherical BS and parameterization}
\begin{figure*}[htbp]
  \centering
  \includegraphics[width=\textwidth]{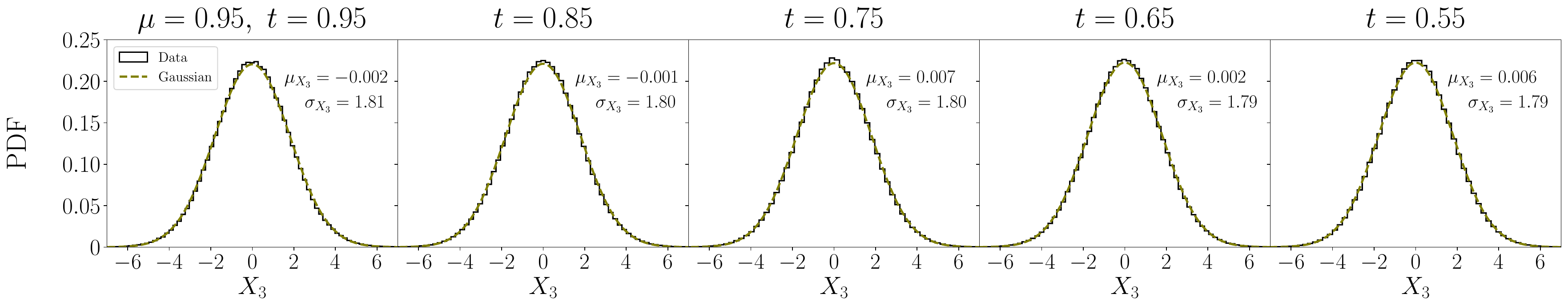}
  \vspace{4mm} 
  \includegraphics[width=\textwidth]{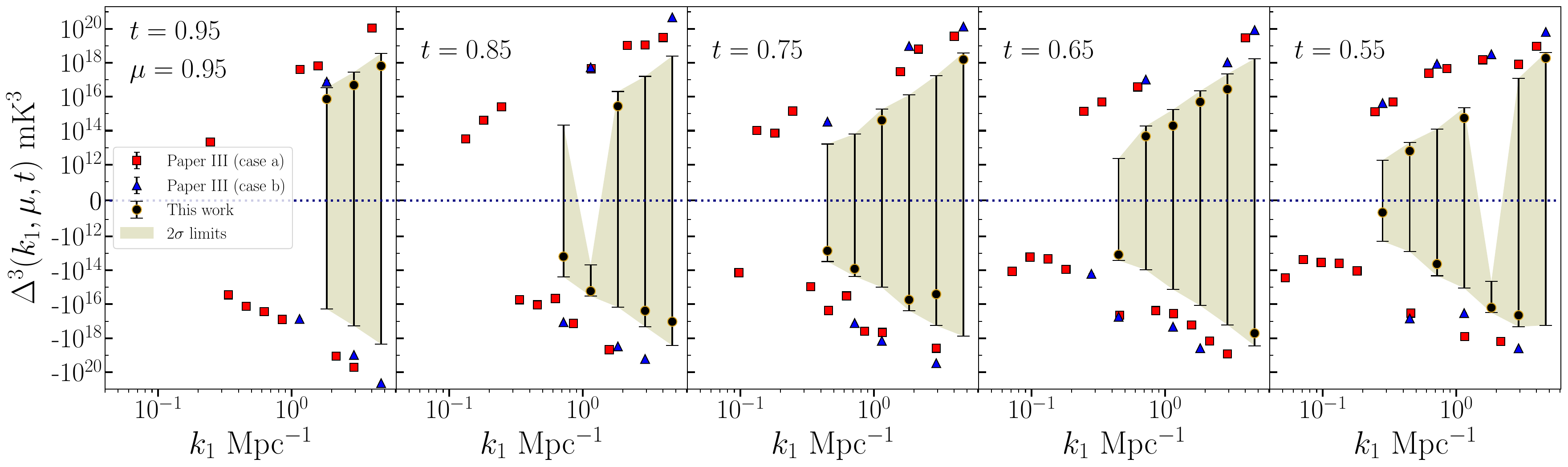}
  \caption{ The top panels show the histogram of variable $X_3$ for a set of near-linear triangles $(\mu=0.95)$, with varying $t$. The $X_3$ statistics quantify how much the BS estimates exceed r.m.s. of noise ${\delta{B}_N}$. The $\mu_{X_3}$ and $\sigma_{X_3}$ denote the mean and standard deviation of the $X_3$ distribution, respectively. The green solid curve represents the best-fit Gaussian. The bottom panels show mean cube brightness temperature fluctuations $\Delta^3$. The black markers with $2\sigma$ error bars show the measurements obtained in the present work (after SCF). The green shaded band indicates the constrained range for the EoR signal, with its lower and upper boundaries marking the derived lower and upper limits. The red squares show the upper limit from \citetalias{Gill_2025_mwa1} (before SCF), computed for triangles whose three sides lie outside the foreground wedge (scenario A3 in Fig.~6 of \citetalias{Gill_2025_mwa1}). The blue triangles also display the results before SCF, but consider the same mode selection used to obtain the black markers.}  
  \label{fig:3dbs5}
\end{figure*}

Finally, we spherically bin the 3D BS $B(k_{1\perp}, k_{2\perp}, k_{3\perp}, k_{1\parallel}, k_{2\parallel})$ using the modes within the region $(k_{1 \perp},k_{2 \perp},k_{3 \perp})\leq 0.026~{\rm Mpc}^{-1}$ and $(k_{1\parallel},k_{2\parallel},k_{3\parallel})>0.135~{\rm Mpc}^{-1}$, and evaluate the spherical BS $\bar{B}(k_1,k_2,k_3)$, where $k_1=\sqrt{k_{1\perp}^2 + k_{1\parallel}^2}$, etc.  Here, $(k_1,k_2,k_3)$ denotes the lengths of the three sides of a triangle  $(k_1\ge k_2 \ge k_3)$, which uniquely specifies its shape and size. In this work, we adopt a convenient parametrization \citep{bharad2020},  which uses the largest side $k_1$ to quantify the size and the two dimensionless parameters 
\begin{equation}
\mu =- \frac{\kk_1 \cdot \kk_2}{k_1 k_2}, \hspace{0.5cm} \, t = \frac{k_2}{k_1} \,,
\label{eq:shape}
\end{equation}
to quantify the shape. 
The values of the parameters $(\mu,t)$ are bounded in the range  
\begin{equation}
    0.5 \leq \mu, t \leq 1 \hspace{0.5cm}  {\rm and} \hspace{0.5cm}  2\mu t \geq 1 \,.
    \label{eq:prange}
\end{equation}
The Appendix of \citetalias{Gill_2025_mabs} and several earlier works present the detailed description of this parameterization, and we do not repeat it here. 
 
The estimated values of the spherical BS $\bar{B}(k_1, \mu, t)$ do not sample the $(k_1,\mu, t)$ space uniformly. In order to obtain a more uniform coverage and also enhance the signal-to-noise ratio (SNR), we perform an additional binning in the $(k_1, \mu, t)$ space. The $k_1$ range $[0.135~{\rm Mpc}^{-1}, 9.25 ~{\rm Mpc}^{-1}]$, is divided into 9 equal logarithmically spaced bins. Note that the number of triangles for $k_1\gtrsim 4.7~{\rm Mpc}^{-1}$ is very small; consequently, the last bin is excluded from the analysis. The bin-averaged $k_1$ values for which the BS estimates are reported therefore span the interval $[0.281~{\rm Mpc}^{-1}, 4.699 ~{\rm Mpc}^{-1}]$. The $\mu$ and $t$ ranges $[0.5,1]$ are each divided into $5$ linearly spaced bins. 
For each ($k_1,\mu,t$) bin, we compute a weighted average of the estimated spherical BS using the number of triangles as weights. 
We then have uniformly spaced estimates of $\bar{B}(k_1, \mu, t)$. 
Because of our selected  region in $(k_{1\parallel},k_{2\parallel},k_{3\parallel})$ and $(k_{1\perp},k_{2\perp},k_{3\perp})$ space, not all triangle shapes are available. We see 
(Fig.~\ref{fig:wedge}) that the $(\kk_1,\kk_2,\kk_3)$ modes are primarily dominated by 
$(k_{1\parallel},k_{2\parallel},k_{3\parallel})$, whereby the resultant triangle shapes are nearly all linear ($\mu=0.95$), for which the three sides are aligned along $k_{\parallel}$. 
In the subsequent analysis, we present results only for linear triangles, i.e., the bins with $\mu=0.95$.

\subsection{Statistics of the BS estimates}


We begin by assessing the statistics of the estimated $B(k_{1\perp}, k_{2\perp}, k_{3\perp}, k_{1\parallel}, k_{2\parallel})$.  \citet{Pal2021} have introduced a quantity $X$ to analyze the statistics of the estimated cylindrical PS. This is defined as the ratio of the estimated value $P(k_{\perp},k_{\parallel})$  to the expected statistical fluctuations due to the system noise $\delta {P}_N(k_{\perp},k_{\parallel})$. The values of $X$ are expected to be symmetrically distributed with mean $\mu_X \approx 0$ and variance $\sigma^2_X \approx 1$ if the estimated values of the cylindrical PS are consistent with predicted system noise.  Deviations from $\mu_X \approx 0$, or an asymmetric distribution,  could indicate the presence of systematics, such as residual foregrounds, etc. Deviations from $\sigma_X \approx 1$ could indicate that the amplitude of the system noise, which depends on the sky temperature $T_{\rm sky}$, has been incorrectly predicted 
\citep{Pal2022, Elahi2023, Elahi2023b,  Elahi_2024, Elahi_missing, chatterjee_2024}.

In the present work, we introduce  the quantity 
\begin{equation}
    X_3 = \dfrac{{B}(k_{1\perp}, k_{2\perp}, k_{3\perp}, k_{1\parallel}, k_{2\parallel})}{\delta {B}_N(k_{1\perp}, k_{2\perp}, k_{3\perp}, k_{1\parallel}, k_{2\parallel})} \,.
\end{equation}
to quantify the statistics of the estimated cylindrical BS. Similar to $X$,   $X_3$ is the ratio of the estimated value to the expected statistical fluctuations due to system noise.

We have a $X_3$ value for every estimate of the cylindrical BS. The estimated spherical BS $B(k_1,\mu,t)$ combines the contribution from several different estimates of the cylindrical BS $B(k_{1\perp}, k_{2\perp}, k_{3\perp}, k_{1\parallel}, k_{2\parallel})$.  Here, for each $(\mu,t)$ bin, we consider   $\{{X_3}\}_{\mu,t}$, which is the set of $X_3$ values corresponding to all the different cylindrical BS estimates that correspond to this particular $(\mu,t)$ bin, irrespective of the value of $k_1$.   We then analyze (top row of Fig.~\ref{fig:3dbs5}) the distribution of $X_3$ for every $(\mu,t)$ bin, and compute the mean $\mu_{X_3}$ and standard deviation $\sigma_{X_3}$. We expect $X_3$ to have a symmetric distribution with  $\mu_{X_3} \approx 0$ and $\sigma_{X_3} \approx 1$, if the estimated cylindrical BS values are consistent with the statistical fluctuations predicted due to system noise. We expect any systematics like residual foregrounds to make the distribution of $X_3$ asymmetric with  $\mid \mu_{X_3} \mid >0$.  Furthermore, a symmetric distribution with $\sigma_{X_3}>1$ implies that the actual uncertainties in the estimated BS are larger than those predicted from the noise-only simulations, possibly due to $T_{sky}$ being underestimated or due to the presence of unknown factors that contribute to the statistical fluctuations.

The top row of Fig.~\ref{fig:3dbs5} shows the probability density function (PDF) of  $X_3$ for different possible triangle shapes, with $\mu=0.95$ fixed and $t$ varying from $0.95$ (squeezed configuration) to $0.55$ (stretched configuration). We observe that the histograms of $X_3$, shown in black lines, are symmetric with the mean remaining very close to zero $\mu_{X_3}\in[-0.002,0.007]$ and the standard deviation $\sigma_{X_3}\in[1.79,1.81]$. For all cases, it follows a Gaussian distribution. The green solid lines in each panel show the fit to the Gaussian distribution.

The values $\sigma_{X_3} > 1$ indicate that the measured BS has fluctuations that are larger than those predicted from the system noise only. Such excess uncertainties have previously been found in power spectrum analysis from the same data \citep{chatterjee_2024, Elahi_missing}, and also in other datasets at low redshifts \citep{Pal2022,Elahi_2024}. The fact that the sky temperature $T_{\rm sky}$ actually varies for different parts of the sky, whereas it has been held fixed for predicting the system noise contribution, can possibly account for  $\sigma_{X_3} > 1$. It is also possible that additional, unknown sources of fluctuation are present in the data. Either way, $\sigma_{X_3}$ has values that are close to unity, which indicates that these effects are small.  Here, we account for these effects  by scaling the noise predictions, $\delta{B}_N$, by $\sigma_{X_3}$. In the subsequent analysis, we use $\delta{B}_N^{\rm True}=\sigma_{X_3} \times \delta{B}_N$ to predict the statistical fluctuations expected in the estimated BS. We note that these results indicate that the cylindrical BS estimates considered here are largely noise-dominated, with the residual foreground contamination being subdominant.


\subsection{BS estimates for various triangle configurations}

The bottom row of Fig.~\ref{fig:3dbs5} shows the mean cube brightness temperature fluctuations, defined as
\begin{equation}
    \Delta^3(k_1,\mu,t)=\dfrac  {k_1^6\,  \bar{B}(k_1,\mu,t)} {(2\pi^2)^{2}} \,.
\label{eq:mcb}
\end{equation}

The black circles show the estimates obtained in this study after applying the SCF. The corresponding error bars show the $\pm 2\sigma$ statistical fluctuations, where $\sigma={(2\pi^2)^{-2}}\,k_1^6\,\delta\bar{B}_N^{\rm True}$. Note that, unlike the PS, the BS can have both positive and negative values.  For comparison, we also show the results before the SCF from \citetalias{Gill_2025_mwa1}  using red squares. This considers all the triangles for which the three sides lie outside the foreground wedge, i.e., $k_{i\parallel}>[k_{i\parallel}]_H$ with  $i=1,2,3$ (scenario A3 in \citetalias{Gill_2025_mwa1}), which we denote as ``case a'' here. The blue triangles show the results before the SCF, considering exactly the same set of triangles as used in this work i.e. $(k_{1 \perp},k_{2 \perp},k_{3 \perp})\leq 0.026~{\rm Mpc}^{-1}$ and $(k_{1\parallel},k_{2\parallel},k_{3\parallel})>0.135~{\rm Mpc}^{-1}$, which we denote as ``case b'' here. As expected, case a spans a wider range of $k_1$ because there is a large variety of triangles in this case.  Considering the results before the SCF, in both cases, the BS has both positive and negative values. The magnitudes of these values are much larger than those after the SCF. Furthermore, the results before the SCF are all much larger in magnitude compared to the predicted $2 \sigma$ error bars, so much so, that the error bars are not visible in the plots. We interpret the estimated spherical BS before the SCF as arising from foregrounds.

Considering the results after SCF, we see that the $k_1$ range increases from left to right across the panels.  We have the smallest range of $k_1$ $\in[1.837,4.699]~{\rm Mpc}^{-1}$ for squeezed triangle ($\mu=0.95,t=0.95$), and the largest range of $k_1$ $\in[0.281,4.699]~{\rm Mpc}^{-1}$ for stretched triangle ($\mu=0.95,t=0.55$). 
 After the SCF, the $B(k_1,\mu,t)$ estimates have both positive and negative values. In all cases, there is a substantial drop in the value of $\mid B(k_1,\mu,t)\mid$   after the SCF as compared to the scenario before the SCF. For nearly all the estimated values, $B(k_1,\mu,t)$ is much smaller than the predicted $2 \sigma$ error bars, and the results are consistent with $\pm 2 \sigma$ statistical fluctuations around zero. There are two estimates of $B(k_1,\mu,t)$ that are not consistent with $2 \sigma$  statistical fluctuations around zero. Both of these estimates have a negative value of $B(k_1,\mu,t)$  with $k_1$ in the range $1-2 \, {\rm Mpc}^{-1}$. We note that both of the estimated values are well within $\pm 3 \sigma$, and it is consequently quite reasonable to also interpret these as being consistent with the predicted statistical fluctuations. 
 
 In summary, the results in the top row and bottom row of  Fig.~\ref{fig:3dbs5} together indicate that all the estimates of the spherical BS presented here are broadly consistent with the predicted statistical fluctuations. There is no evidence for significant foreground contamination.





\subsection{Constraints on the EoR signal}

\begin{figure*}[htbp]
  \centering
  \begin{minipage}[b]{0.49\textwidth}
    \centering
    \includegraphics[width=\linewidth]{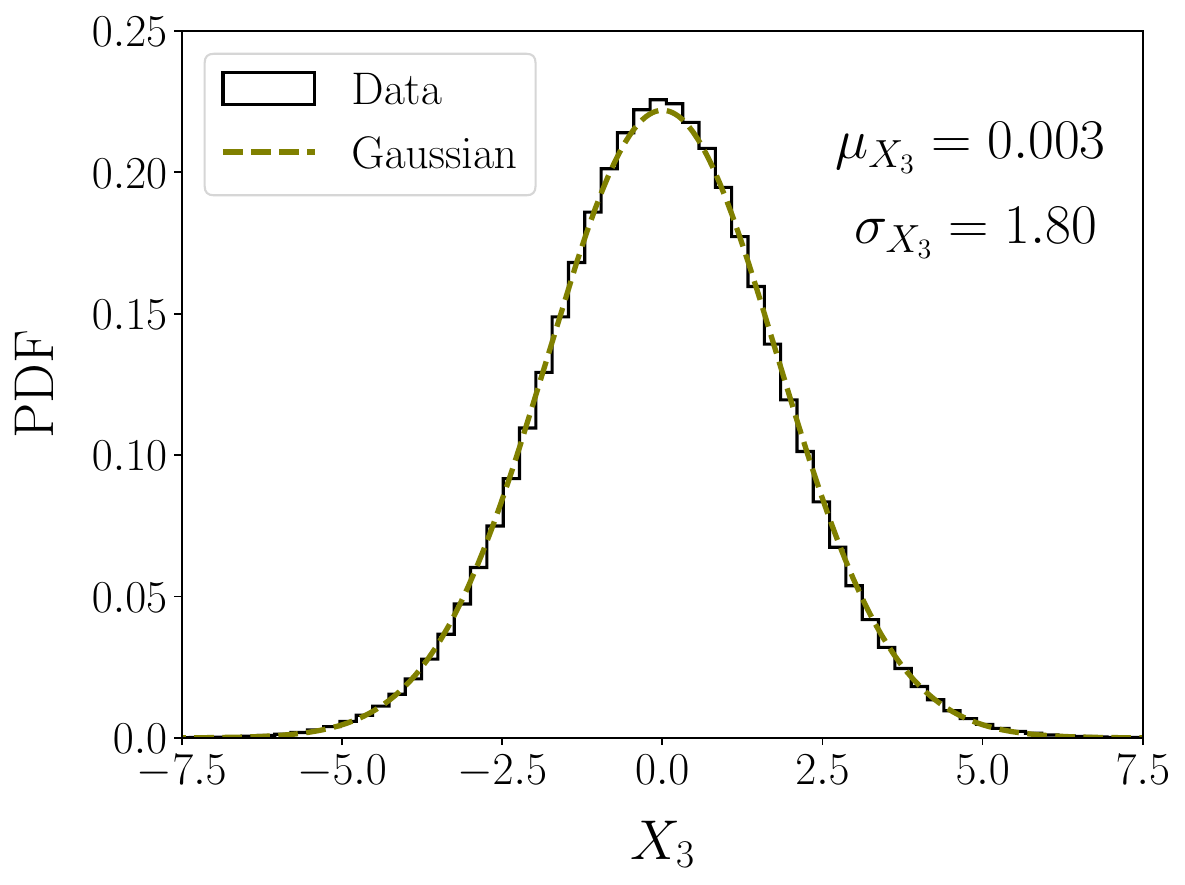}
  \end{minipage}\hfill
  \begin{minipage}[b]{0.49\textwidth}
    \centering
    \includegraphics[width=\linewidth]{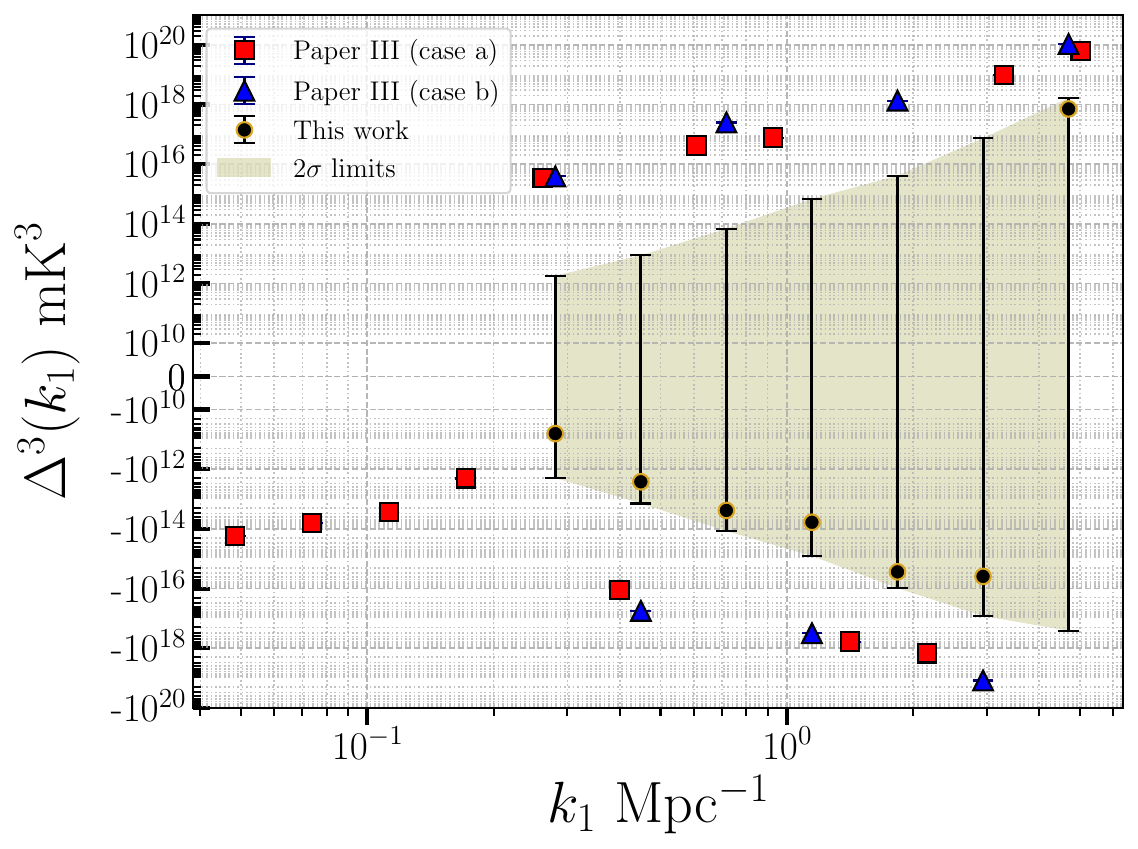}
  \end{minipage}
  \caption{Combining all triangles of Figure \ref{fig:3dbs5}. Left: $X_3$ statistics after assimilating all the BS estimates for selected modes. The green line shows the best-fit Gaussian curve. Right: The black markers with $2\sigma$ error bars show the measurements of $\Delta^3$ after combining all triangle shapes. The green-shaded band marks the constrained range for the EoR signal, with its lower and upper boundaries indicating the derived lower and upper limits. The red squares show the upper limits obtained before SCF in \citetalias{Gill_2025_mwa1} (scenario A3 in Fig.~6 of \citetalias{Gill_2025_mwa1}). The blue triangles also display the results before SCF, but use the same mode selection as for the black markers.}
  \label{fig:3dbs1}
\end{figure*}

\setlength{\tabcolsep}{20pt} 
\renewcommand{\arraystretch}{1.09} 

\begin{table*}[ht]
\centering
\caption{Measurements of mean cube brightness temperature fluctuations $\Delta^3(k_1,\mu,t)$, their corresponding $1\sigma$ systematical uncertainties, signal to noise ratio (SNR), $2\sigma$ lower limit $\Delta^3_{\rm LL}=\Delta^3-2\sigma$, and upper limit $\Delta^3_{\rm UL}=\Delta^3+2\sigma$.}
\label{tab:bispec}
{\setlength{\heavyrulewidth}{1.2pt}%
\begin{tabular}{l c c c c c}
\toprule
{$k_1$} & {$\Delta^3$} & {$1\sigma$} & {SNR} & {$\Delta^3_{\rm LL}$} & {$\Delta^3_{\rm UL}$} \\
{[Mpc$^{-1}$]} & \multicolumn{1}{c}{[mK$^{3}$]} & \multicolumn{1}{c}{[mK$^{3}$]} & & \multicolumn{1}{c}{[mK$^{3}$]} & \multicolumn{1}{c}{[mK$^{3}$]} \\
\midrule
0.281 & $-(4.00\times10^{3})^3$ & $(9.81\times10^{3})^3$ & $0.068$ & $-(1.25\times10^{4})^3$ & $(1.22\times10^{4})^3$ \\
0.449 & $-(1.38\times10^{4})^3$ & $(1.79\times10^{4})^3$ & $0.460$ & $-(2.41\times10^{4})^3$ & $(2.06\times10^{4})^3$ \\
0.718 & $-(2.88\times10^{4})^3$ & $(3.60\times10^{4})^3$ & $0.513$ & $-(4.90\times10^{4})^3$ & $(4.11\times10^{4})^3$ \\
1.148 & $-(3.91\times10^{4})^3$ & $(7.20\times10^{4})^3$ & $0.160$ & $-(9.31\times10^{4})^3$ & $(8.82\times10^{4})^3$ \\
1.837 & $-(1.40\times10^{5})^3$ & $(1.51\times10^{5})^3$ & $0.796$ & $-(2.13\times10^{5})^3$ & $(1.61\times10^{5})^3$ \\
2.938 & $-(1.57\times10^{5})^3$ & $(3.39\times10^{5})^3$ & $0.099$ & $-(4.35\times10^{5})^3$ & $(4.20\times10^{5})^3$ \\
4.699 &  $(8.93\times10^{5})^3$ & $(7.86\times10^{5})^3$ & $1.468$ & $-(6.37\times10^{5})^3$ & $(1.19\times10^{6})^3$ \\
\bottomrule
\end{tabular}%
} 
\end{table*}

In order to tighten the constraints on the EoR 21 cm BS, we have combined the estimates of $B(k_1,\mu,t)$ corresponding to different shapes $(\mu,t)$ to obtain $B(k_1)$. 
Fig.~\ref{fig:3dbs1} presents the results averaged over all triangle shapes. The left panel shows the PDF of $X_3$, which essentially incorporates all $N_s=18 \times 10^6$ estimates of $B(k_{1\perp}, k_{2\perp}, k_{3\perp}, k_{1\parallel}, k_{2\parallel})$ available from the data. This corresponds to combining all the data in the different panels of the top row of Fig.~\ref{fig:3dbs5}.  
The resulting distribution has a mean $\mu_{X_3}=0.003$ and a standard deviation $\sigma_{X_3}=1.8$. We see that the data is well fit by a Gaussian distribution (green curve). We now consider the hypothesis that the measured value of $\mu_{X_3}=0.003$ is consistent with $\mu_{X_3}=0$, within the predicted statistical fluctuations due to our finite sample size $N_s$. These fluctuations are predicted to have a standard deviation of $\sigma_{X_3}/\sqrt{N_s}=4 \times 10^{-4}$, which is much smaller than the measured $\mu_{X_3}$.  This rules out our hypothesis and indicates that the measurements are not consistent with the statistical fluctuations expected from noise alone. There is some residual foreground that is responsible for the excess value of  $\mu_{X_3}$. 

The right panel of Fig.~\ref{fig:3dbs1}  shows $\Delta^3(k_1)$ obtained after averaging over triangles of all shapes. The black circles show the results after SCF. We see that, compared to Fig.~\ref{fig:3dbs5},  the averaging reduces the $2 \sigma$ error bars for the estimated values.  The estimated values are all well within the $\pm 2 \sigma$ error bars. The estimated values  $\Delta^3(k_1)$,  along with the $2 \sigma$ upper limits  $\Delta^3_{\rm UL}(k_1)$, and   the $2 \sigma$ lower limits  $\Delta^3_{\rm LL}(k_1)$,  are all  tabulated in Table \ref{tab:bispec}, which also shows ${\rm SNR} = \mid\Delta^3\mid/\sigma$. We see that the estimated values are negative and ${\rm SNR}<1$ for all the $k_1$ bins,  except the largest bin at $k_1=4.669~{\rm Mpc}^{-1}$ where $\Delta^3(k_1)$ is positive and   ${\rm SNR = 1.468}$. 
We find that excluding the largest  $k_1$ bin leaves the value of  $\sigma_{X_3}=1.8$ unchanged. However, we have $\mu_{X_3}=-9 \times 10^{-6}$, which is considerably smaller compared to the earlier value of $\mu_{X_3}=0.003$. Considering $\sigma_{X_3}/\sqrt{N_s}=7 \times 10^{-4}$  for $N_s=6 \times 10^6$, we see that now the value of $\mu_{X_3}$ is consistent with $\mu_{X_3}=0$.  
This tells us that the estimates of  $B(k_{1\perp}, k_{2\perp}, k_{3\perp}, k_{1\parallel}, k_{2\parallel})$ within $k_1 < 3 \, {\rm Mpc}^{-1}$, are consistent with zero within the statistical fluctuations expected from the system noise alone. 
In other words, considering the estimated values $\Delta^3(k_1)$,  the $2 \sigma$ upper limits  $\Delta^3_{\rm UL}(k_1)$, and the $2 \sigma$ lower limits  $\Delta^3_{\rm LL}(k_1)$ presented in Fig.~\ref{fig:3dbs1} and Table \ref{tab:bispec}, at the level of sensitivity achieved here, we expect these to be free of foreground contamination at   $k_1 < 3 \, {\rm Mpc}^{-1}$.

 The most stringent lower limit $\Delta^3_{\rm LL}=-(1.25\times 10^4)^3~{\rm mK}^3$ and upper limit $\Delta^3_{\rm UL}=(1.22\times 10^4)^3~{\rm mK}^3$ that we obtain here  are at smallest $k$ bin  with $k_1=0.281~{\rm Mpc}^{-1}$. The $z\approx 8$ EoR 21 cm signal is predicted to be $\Delta^3\approx -10^3~{\rm mK}^3$ at $k_1\approx 0.29~{\rm Mpc}^{-1}$ \citep{gill_eormulti}, which is approximately nine orders of magnitude smaller than the limits obtained here.



\section{Summary and Conclusions}
\label{sec:sum}

The 21 cm signal from the EoR is highly non-Gaussian due to the emergence and growth of the ionized bubbles \citep{Bharadwaj2005, Majumdar_2018, gill_eormulti}. The PS is incapable of accessing the plethora of information encoded in the non-Gaussianity, and we need to consider the higher-order statistics to exploit the maximum information present in the observational data. The BS is the lowest-order statistic sensitive to non-Gaussianity and also captures the mode coupling and topology of the underlying \Hi~distribution. While the utility of the BS in gaining valuable insights about the EoR has been widely explored through simulations, attempts to measure it using actual observational data are scarce. \citet{trott2019} provided the measurements of the BS for a few triangle configurations using the MWA Phase II data. In \citetalias{Gill_2025_mwa1}, we reported the BS measurements across all possible triangle configurations using a drift-scan MWA data at $z=8.2$. However, the BS estimates were found to be foreground-dominated, even after attempting to mitigate the foregrounds using several ``foreground avoidance'' scenarios.

In this work, we continue the efforts to extract the EoR 21 cm BS from radio-interferometric data used in \citetalias{Gill_2025_mwa1}. Here, we employ the SCF \citep{Elahi_missing}  to mitigate foreground contamination from the BS measurements. It is a frequency-based, powerful method that exploits the fact that foregrounds are spectrally smoother than the \Hi~21 cm signal. We proceed by gridding the observed visibility data, identifying the smooth component from these gridded visibilities on $2~{\rm MHz}$ scale, and using it to filter out the residual visibilities. We use these residual visibilities as our data and employ them to estimate the MABS and the 3D 21 cm BS, utilizing the \texttt{AMBER} estimator (\citetalias{Gill_2025_mabs}). We have validated the full pipeline using a simulated 21 cm signal, and show that we are able to recover the input BS above the smoothing scale $k_{i\parallel}>0.135~{\rm Mpc}^{-1}$, where $i=1,2,3$.

The foreground contamination is significantly reduced in the estimated MABS $B_A(\ell_1,\ell_2,\ell_3,\Delta\nu_1,\Delta\nu_2)$ (Fig. \ref{fig:mabs}) and the 3D cylindrical BS $B(k_{1\perp}, k_{2\perp}, k_{3\perp}, k_{1\parallel}, k_{2\parallel})$ (Fig. \ref{fig:cylbs}) after the SCF. We see the reduction in amplitude by $3-4$ orders of magnitude. It also reduces the periodic square grid pattern in the $B(k_{1\parallel},k_{2\parallel})$, which was present before the SCF. Consequently, the SCF reduces the foreground leakage in the EoR window, and the periodic spikes along the $k_{1\parallel}$ caused by missing frequency channels are mitigated significantly. We have a quite cleaner region in the $B(k_{1\perp},k_{1\parallel})$ (Fig.~\ref{fig:wedge}) for $k_{1\perp}\le0.026~{\rm Mpc}^{-1}$.

We use the $B(k_{1\perp}, k_{2\perp}, k_{3\perp}, k_{1\parallel}, k_{2\parallel})$ measurements within the region $(k_{1\parallel},k_{2\parallel},k_{3\parallel})>0.135~{\rm Mpc}^{-1}$ and $(k_{1\perp},k_{2\perp},k_{3\perp})\le0.026~{\rm Mpc}^{-1}$ to evaluate the 3D spherical BS $B(k_1,\mu,t)$, where $k_1$ and $(\mu,t)$ respectively quantify the size and shape of the triangle $(k_1,k_2,k_3)$. The $(k_1,\mu,t)$ space is further binned into $(9,5,5)$ bins (Fig.~\ref{fig:3dbs5}). The resulting estimates are obtained for linear triangles $\mu=0.95$, with the squeezed triangles ($t=0.95$) spanning smallest range of $k_1\in[1.837,4.699]~{\rm Mpc}^{-1}$ and the stretched triangles ($t=0.55$) spanning widest range of $k_1\in[0.281,4.699]~{\rm Mpc}^{-1}$. The amplitude of the BS is significantly reduced after the SCF for all cases, and the current estimates are dominated by system noise. We analyze the $X_3$ statistics to determine how much the BS estimates exceed the statistical error due to the system noise. For all triangle shapes, the PDF of $X_3$ follows Gaussian distribution with mean remaining very close to zero $\mu_{X_3}\in[-0.002,0.007]$ and the standard deviation $\sigma_{X_3}\in[1.79,1.81]$. The $\sigma_{X_3}>1$ indicates that the statistical fluctuations predicted from the system noise-only simulations are underestimated, and we account for this underestimation by multiplying the uncertainties by $\sigma_{X_3}$.

We combine the estimates corresponding to all triangle shapes ($\mu,t$) to place the constraints on the EoR signal (Fig.~\ref{fig:3dbs1}). The combined PDF of ${X_3}$ for all the estimates has mean $\mu_{X_3}=0.003$, $\sigma_{X_3}=1.8$ and follows the Gaussian distribution. The BS estimates are consistent with the statistical fluctuations due to the system noise, provided we do not exclude the largest $k_1$ bin and impose $k_1 < 3 \, {\rm Mpc}^{-1}$ whereby $\mu_{X_3}$ drops to $9 \times 10^{-6}$. 
Table \ref{tab:bispec} tabulates the BS estimates, $1\sigma$ error bars, SNR, $2\sigma$ lower limit $\Delta_{\rm LL}^3$, and $2\sigma$ upper limit $\Delta_{\rm UL}^3$ on the EoR signal. The best lower limit $\Delta^3_{\rm LL}=-(1.25\times 10^4)^3~{\rm mK}^3$ and upper limit $\Delta^3_{\rm UL}=(1.22\times 10^4)^3~{\rm mK}^3$ are obtained at $k_1=0.281~{\rm Mpc}^{-1}$. The $z\approx 8$ EoR BS signal is predicted to be $\Delta^3\approx -10^3~{\rm mK}^3$ at $k_1\approx 0.29~{\rm Mpc}^{-1}$ \citep{gill_eormulti}.

In summary, while the SCF has significantly reduced the foreground contribution in the data, the BS estimates are still quite larger than the expected EoR signal and are currently dominated by system noise. More hours of observations will help us reduce the system noise contribution and improve the lower and upper limits on the EoR signal. In future work, we will include the data corresponding to all pointing centers of the MWA Phase II drift scan data and place tighter constraints on the EoR 21 cm  BS.

\section*{Acknowledgments}
We acknowledge the computing facilities provided by the Department of Physics at IIT Kharagpur. We acknowledge the National Supercomputing Mission (NSM) for providing computing resources of ‘PARAM Shakti’ at IIT Kharagpur, which is implemented by C-DAC and supported by the Ministry of Electronics and Information Technology (MeitY) and Department of Science and Technology (DST), Government of India.

\section*{Data Availability}

The data sets were derived from sources in the public domain
(the MWA Data Archive: project ID G0031) available at \href{https://asvo.mwatelescope.org/}{MWA ASVO Portal}. 


\bibliography{main}{}

\begin{thebibliography}{}
\expandafter\ifx\csname natexlab\endcsname\relax\def\natexlab#1{#1}\fi
\providecommand{\url}[1]{\href{#1}{#1}}
\providecommand{\dodoi}[1]{doi:~\href{http://doi.org/#1}{\nolinkurl{#1}}}
\providecommand{\doeprint}[1]{\href{http://ascl.net/#1}{\nolinkurl{http://ascl.net/#1}}}
\providecommand{\doarXiv}[1]{\href{https://arxiv.org/abs/#1}{\nolinkurl{https://arxiv.org/abs/#1}}}

\bibitem[{{Bharadwaj} {et~al.}(2020){Bharadwaj}, {Mazumdar}, \& {Sarkar}}]{bharad2020}
{Bharadwaj}, S., {Mazumdar}, A., \& {Sarkar}, D. 2020, \mnras, 493, 594, \dodoi{10.1093/mnras/staa279}

\bibitem[{Bharadwaj \& Pandey(2005)}]{BP2005}
Bharadwaj, S., \& Pandey, S.~K. 2005, Mon. Not. Roy. Astron. Soc., 358, 968, \dodoi{10.1111/j.1365-2966.2005.08836.x}

\bibitem[{{Bharadwaj} \& {Pandey}(2005)}]{Bharadwaj2005}
{Bharadwaj}, S., \& {Pandey}, S.~K. 2005, \mnras, 358, 968, \dodoi{10.1111/j.1365-2966.2005.08836.x}

\bibitem[{{Chatterjee} {et~al.}(2024){Chatterjee}, {Elahi}, {Bharadwaj}, {Sarkar}, {Choudhuri}, {Sethi}, \& {Patwa}}]{chatterjee_2024}
{Chatterjee}, S., {Elahi}, K. M.~A., {Bharadwaj}, S., {et~al.} 2024, \pasa, 41, e077, \dodoi{10.1017/pasa.2024.45}

\bibitem[{{DeBoer} {et~al.}(2017){DeBoer}, {Parsons}, {Aguirre}, {Alexander}, {Ali}, {Beardsley}, {Bernardi}, {Bowman}, {Bradley}, {Carilli}, {Cheng}, {de Lera Acedo}, {Dillon}, {Ewall-Wice}, {Fadana}, {Fagnoni}, {Fritz}, {Furlanetto}, {Glendenning}, {Greig}, {Grobbelaar}, {Hazelton}, {Hewitt}, {Hickish}, {Jacobs}, {Julius}, {Kariseb}, {Kohn}, {Lekalake}, {Liu}, {Loots}, {MacMahon}, {Malan}, {Malgas}, {Maree}, {Martinot}, {Mathison}, {Matsetela}, {Mesinger}, {Morales}, {Neben}, {Patra}, {Pieterse}, {Pober}, {Razavi-Ghods}, {Ringuette}, {Robnett}, {Rosie}, {Sell}, {Smith}, {Syce}, {Tegmark}, {Thyagarajan}, {Williams}, \& {Zheng}}]{DeBoer_2017}
{DeBoer}, D.~R., {Parsons}, A.~R., {Aguirre}, J.~E., {et~al.} 2017, \pasp, 129, 045001, \dodoi{10.1088/1538-3873/129/974/045001}

\bibitem[{{Elahi} {et~al.}(2025){Elahi}, {Bharadwaj}, {Chatterjee}, {Sarkar}, {Choudhuri}, {Sethi}, \& {Patwa}}]{Elahi_missing}
{Elahi}, K. M.~A., {Bharadwaj}, S., {Chatterjee}, S., {et~al.} 2025, \mnras, 540, 2745, \dodoi{10.1093/mnras/staf896}

\bibitem[{{Elahi} {et~al.}(2023{\natexlab{a}}){Elahi}, {Bharadwaj}, {Ghosh}, {Pal}, {Ali}, {Choudhuri}, {Chakraborty}, {Datta}, {Roy}, {Choudhury}, \& {Dutta}}]{Elahi2023}
{Elahi}, K. M.~A., {Bharadwaj}, S., {Ghosh}, A., {et~al.} 2023{\natexlab{a}}, \mnras, 520, 2094, \dodoi{10.1093/mnras/stad191}

\bibitem[{{Elahi} {et~al.}(2023{\natexlab{b}}){Elahi}, {Bharadwaj}, {Pal}, {Ghosh}, {Ali}, {Choudhuri}, {Chakraborty}, {Datta}, {Roy}, {Choudhury}, \& {Dutta}}]{Elahi2023b}
{Elahi}, K. M.~A., {Bharadwaj}, S., {Pal}, S., {et~al.} 2023{\natexlab{b}}, \mnras, 525, 3439, \dodoi{10.1093/mnras/stad2495}

\bibitem[{{Elahi} {et~al.}(2024){Elahi}, {Bharadwaj}, {Pal}, {Ghosh}, {Ali}, {Choudhuri}, {Chakraborty}, {Datta}, {Roy}, {Choudhury}, \& {Dutta}}]{Elahi_2024}
---. 2024, \mnras, 529, 3372, \dodoi{10.1093/mnras/stae740}

\bibitem[{Gill \& Bharadwaj(2025)}]{Gill_2025_mabs}
Gill, S.~S., \& Bharadwaj, S. 2025

\bibitem[{{Gill} {et~al.}(2025{\natexlab{a}}){Gill}, {Bharadwaj}, {Ali}, \& {Elahi}}]{Gill_2024_2d3vc}
{Gill}, S.~S., {Bharadwaj}, S., {Ali}, S.~S., \& {Elahi}, K. M.~A. 2025{\natexlab{a}}, \apj, 979, 25, \dodoi{10.3847/1538-4357/ad9b20}

\bibitem[{{Gill} {et~al.}(2025{\natexlab{b}}){Gill}, {Bharadwaj}, {Elahi}, {Sethi}, \& {Patwa}}]{Gill_2025_mwa1}
{Gill}, S.~S., {Bharadwaj}, S., {Elahi}, K. M.~A., {Sethi}, S.~K., \& {Patwa}, A.~K. 2025{\natexlab{b}}, \apj, 993, 56, \dodoi{10.3847/1538-4357/ae0463}

\bibitem[{{Gill} {et~al.}(2024){Gill}, {Pramanick}, {Bharadwaj}, {Shaw}, \& {Majumdar}}]{gill_eormulti}
{Gill}, S.~S., {Pramanick}, S., {Bharadwaj}, S., {Shaw}, A.~K., \& {Majumdar}, S. 2024, \mnras, 527, 1135, \dodoi{10.1093/mnras/stad3273}

\bibitem[{{Gupta} {et~al.}(2017){Gupta}, {Ajithkumar}, {Kale}, {Nayak}, {Sabhapathy}, {Sureshkumar}, {Swami}, {Chengalur}, {Ghosh}, {Ishwara-Chandra}, {Joshi}, {Kanekar}, {Lal}, \& {Roy}}]{Gupta_2017}
{Gupta}, Y., {Ajithkumar}, B., {Kale}, H.~S., {et~al.} 2017, Current Science, 113, 707, \dodoi{10.18520/cs/v113/i04/707-714}

\bibitem[{{HERA Collaboration} {et~al.}(2023){HERA Collaboration}, {Abdurashidova}, {Adams}, {Aguirre}, {Alexander}, {Ali}, {Baartman}, {Balfour}, {Barkana}, {Beardsley}, {Bernardi}, {Billings}, {Bowman}, {Bradley}, {Breitman}, {Bull}, {Burba}, {Carey}, {Carilli}, {Cheng}, {Choudhuri}, {DeBoer}, {de Lera Acedo}, {Dexter}, {Dillon}, {Ely}, {Ewall-Wice}, {Fagnoni}, {Fialkov}, {Fritz}, {Furlanetto}, {Gale-Sides}, {Garsden}, {Glendenning}, {Gorce}, {Gorthi}, {Greig}, {Grobbelaar}, {Halday}, {Hazelton}, {Heimersheim}, {Hewitt}, {Hickish}, {Jacobs}, {Julius}, {Kern}, {Kerrigan}, {Kittiwisit}, {Kohn}, {Kolopanis}, {Lanman}, {La Plante}, {Lewis}, {Liu}, {Loots}, {Ma}, {MacMahon}, {Malan}, {Malgas}, {Malgas}, {Maree}, {Marero}, {Martinot}, {McBride}, {Mesinger}, {Mirocha}, {Molewa}, {Morales}, {Mosiane}, {Mu{\~n}oz}, {Murray}, {Nagpal}, {Neben}, {Nikolic}, {Nunhokee}, {Nuwegeld}, {Parsons}, {Pascua}, {Patra}, {Pieterse}, {Qin}, {Razavi-Ghods}, {Robnett}, {Rosie}, {Santos}, {Sims}, {Singh}, {Smith}, {Swarts}, {Tan},
  {Thyagarajan}, {Wilensky}, {Williams}, {van Wyngaarden}, \& {Zheng}}]{HERA2023}
{HERA Collaboration}, {Abdurashidova}, Z., {Adams}, T., {et~al.} 2023, \apj, 945, 124, \dodoi{10.3847/1538-4357/acaf50}

\bibitem[{Hutter {et~al.}(2020)Hutter, Watkinson, Seiler, Dayal, Sinha, \& Croton}]{Hutter_2019}
Hutter, A., Watkinson, C.~A., Seiler, J., {et~al.} 2020, MNRAS, 492, 653, \dodoi{10.1093/mnras/stz3139}

\bibitem[{{Kamran} {et~al.}(2022){Kamran}, {Ghara}, {Majumdar}, {Mellema}, {Bharadwaj}, {Pritchard}, {Mondal}, \& {Iliev}}]{kamran_2022}
{Kamran}, M., {Ghara}, R., {Majumdar}, S., {et~al.} 2022, JCAP, 2022, 001, \dodoi{10.1088/1475-7516/2022/11/001}

\bibitem[{{Kamran} {et~al.}(2021){Kamran}, {Majumdar}, {Ghara}, {Mellema}, {Bharadwaj}, {Pritchard}, {Mondal}, \& {Iliev}}]{Kamran_2021b}
{Kamran}, M., {Majumdar}, S., {Ghara}, R., {et~al.} 2021, arXiv e-prints, arXiv:2108.08201, \dodoi{10.48550/arXiv.2108.08201}

\bibitem[{{Koopmans} {et~al.}(2015){Koopmans}, {Pritchard}, {Mellema}, {Aguirre}, {Ahn}, {Barkana}, {van Bemmel}, {Bernardi}, {Bonaldi}, {Briggs}, {de Bruyn}, {Chang}, {Chapman}, {Chen}, {Ciardi}, {Dayal}, {Ferrara}, {Fialkov}, {Fiore}, {Ichiki}, {Illiev}, {Inoue}, {Jelic}, {Jones}, {Lazio}, {Maio}, {Majumdar}, {Mack}, {Mesinger}, {Morales}, {Parsons}, {Pen}, {Santos}, {Schneider}, {Semelin}, {de Souza}, {Subrahmanyan}, {Takeuchi}, {Vedantham}, {Wagg}, {Webster}, {Wyithe}, {Datta}, \& {Trott}}]{Koopmans_2015}
{Koopmans}, L., {Pritchard}, J., {Mellema}, G., {et~al.} 2015, in Advancing Astrophysics with the Square Kilometre Array (AASKA14), 1, \dodoi{10.22323/1.215.0001}

\bibitem[{Majumdar {et~al.}(2020)Majumdar, Kamran, Pritchard, Mondal, Mazumdar, Bharadwaj, \& Mellema}]{Majumdar_2020}
Majumdar, S., Kamran, M., Pritchard, J.~R., {et~al.} 2020, MNRAS, 499, 5090, \dodoi{10.1093/mnras/staa3168}

\bibitem[{Majumdar {et~al.}(2018{\natexlab{a}})Majumdar, Pritchard, Mondal, Watkinson, Bharadwaj, \& Mellema}]{suamnm2018}
Majumdar, S., Pritchard, J.~R., Mondal, R., {et~al.} 2018{\natexlab{a}}, Mon. Not. Roy. Astron. Soc., 476, 4007, \dodoi{10.1093/mnras/sty535}

\bibitem[{Majumdar {et~al.}(2018{\natexlab{b}})Majumdar, Pritchard, Mondal, Watkinson, Bharadwaj, \& Mellema}]{Majumdar_2018}
---. 2018{\natexlab{b}}, MNRAS, 476, 4007, \dodoi{10.1093/mnras/sty535}

\bibitem[{{Mondal} {et~al.}(2015){Mondal}, {Bharadwaj}, {Majumdar}, {Bera}, \& {Acharyya}}]{mondal2015}
{Mondal}, R., {Bharadwaj}, S., {Majumdar}, S., {Bera}, A., \& {Acharyya}, A. 2015, \mnras, 449, L41, \dodoi{10.1093/mnrasl/slv015}

\bibitem[{Murmu {et~al.}(2023)Murmu, Datta, Majumdar, \& Greve}]{Murmu:2023fgv}
Murmu, C.~S., Datta, K.~K., Majumdar, S., \& Greve, T.~R. 2023.
\newblock \doarXiv{2311.17062}

\bibitem[{{Nuttall}(1981)}]{Nuttall_1981}
{Nuttall}, A.~H. 1981, IEEE Transactions on Acoustics Speech and Signal Processing, 29, 84

\bibitem[{Pal {et~al.}(2020)Pal, Bharadwaj, Ghosh, \& Choudhuri}]{Pal2021}
Pal, S., Bharadwaj, S., Ghosh, A., \& Choudhuri, S. 2020, Monthly Notices of the Royal Astronomical Society, 501, 3378, \dodoi{10.1093/mnras/staa3831}

\bibitem[{{Pal} {et~al.}(2022){Pal}, {Elahi}, {Bharadwaj}, {Ali}, {Choudhuri}, {Ghosh}, {Chakraborty}, {Datta}, {Roy}, {Choudhury}, \& {Dutta}}]{Pal2022}
{Pal}, S., {Elahi}, K. M.~A., {Bharadwaj}, S., {et~al.} 2022, \mnras, 516, 2851, \dodoi{10.1093/mnras/stac2419}

\bibitem[{{Patwa} {et~al.}(2021){Patwa}, {Sethi}, \& {Dwarakanath}}]{Patwa_2021}
{Patwa}, A.~K., {Sethi}, S., \& {Dwarakanath}, K.~S. 2021, \mnras, 504, 2062, \dodoi{10.1093/mnras/stab989}

\bibitem[{{Planck Collaboration} {et~al.}(2020){Planck Collaboration}, {Akrami}, {Arroja}, {Ashdown}, {Aumont}, {Baccigalupi}, {Ballardini}, {Banday}, {Barreiro}, {Bartolo}, {Basak}, {Benabed}, {Bernard}, {Bersanelli}, {Bielewicz}, {Bond}, {Borrill}, {Bouchet}, {Bucher}, {Burigana}, {Butler}, {Calabrese}, {Cardoso}, {Casaponsa}, {Challinor}, {Chiang}, {Colombo}, {Combet}, {Crill}, {Cuttaia}, {de Bernardis}, {de Rosa}, {de Zotti}, {Delabrouille}, {Delouis}, {Di Valentino}, {Diego}, {Dor{\'e}}, {Douspis}, {Ducout}, {Dupac}, {Dusini}, {Efstathiou}, {Elsner}, {En{\ss}lin}, {Eriksen}, {Fantaye}, {Fergusson}, {Fernandez-Cobos}, {Finelli}, {Frailis}, {Fraisse}, {Franceschi}, {Frolov}, {Galeotta}, {Galli}, {Ganga}, {G{\'e}nova-Santos}, {Gerbino}, {Gonz{\'a}lez-Nuevo}, {G{\'o}rski}, {Gratton}, {Gruppuso}, {Gudmundsson}, {Hamann}, {Handley}, {Hansen}, {Herranz}, {Hivon}, {Huang}, {Jaffe}, {Jones}, {Jung}, {Keih{\"a}nen}, {Keskitalo}, {Kiiveri}, {Kim}, {Krachmalnicoff}, {Kunz}, {Kurki-Suonio}, {Lamarre}, {Lasenby},
  {Lattanzi}, {Lawrence}, {Le Jeune}, {Levrier}, {Lewis}, {Liguori}, {Lilje}, {Lindholm}, {L{\'o}pez-Caniego}, {Ma}, {Mac{\'\i}as-P{\'e}rez}, {Maggio}, {Maino}, {Mandolesi}, {Marcos-Caballero}, {Maris}, {Martin}, {Mart{\'\i}nez-Gonz{\'a}lez}, {Matarrese}, {Mauri}, {McEwen}, {Meerburg}, {Meinhold}, {Melchiorri}, {Mennella}, {Migliaccio}, {Miville-Desch{\^e}nes}, {Molinari}, {Moneti}, {Montier}, {Morgante}, {Moss}, {M{\"u}nchmeyer}, {Natoli}, {Oppizzi}, {Pagano}, {Paoletti}, {Partridge}, {Patanchon}, {Perrotta}, {Pettorino}, {Piacentini}, {Polenta}, {Puget}, {Rachen}, {Racine}, {Reinecke}, {Remazeilles}, {Renzi}, {Rocha}, {Rubi{\~n}o-Mart{\'\i}n}, {Ruiz-Granados}, {Salvati}, {Savelainen}, {Scott}, {Shellard}, {Shiraishi}, {Sirignano}, {Sirri}, {Smith}, {Spencer}, {Stanco}, {Sunyaev}, {Suur-Uski}, {Tauber}, {Tavagnacco}, {Tenti}, {Toffolatti}, {Tomasi}, {Trombetti}, {Valiviita}, {Van Tent}, {Vielva}, {Villa}, {Vittorio}, {Wandelt}, {Wehus}, {Zacchei}, \& {Zonca}}]{planck2020}
{Planck Collaboration}, {Akrami}, Y., {Arroja}, F., {et~al.} 2020, \aap, 641, A9, \dodoi{10.1051/0004-6361/201935891}

\bibitem[{{Raste} {et~al.}(2024){Raste}, {Kulkarni}, {Watkinson}, {Keating}, \& {Haehnelt}}]{raste_2023}
{Raste}, J., {Kulkarni}, G., {Watkinson}, C.~A., {Keating}, L.~C., \& {Haehnelt}, M.~G. 2024, \mnras, 529, 129, \dodoi{10.1093/mnras/stae492}

\bibitem[{{Sarkar} {et~al.}(2026){Sarkar}, {Elahi}, {Choudhuri}, {Bharadwaj}, {Chatterjee}, {Bhattacharyya}, {Sethi}, \& {Patwa}}]{sarkar_2026}
{Sarkar}, S., {Elahi}, K. M.~A., {Choudhuri}, S., {et~al.} 2026, (in prep.)

\bibitem[{Saxena {et~al.}(2020)Saxena, Majumdar, Kamran, \& Viel}]{Saxena_2020}
Saxena, A., Majumdar, S., Kamran, M., \& Viel, M. 2020, Mon. Not. Roy. Astron. Soc., 497, 2941, \dodoi{10.1093/mnras/staa1768}

\bibitem[{{Shimabukuro} {et~al.}(2016){Shimabukuro}, {Yoshiura}, {Takahashi}, {Yokoyama}, \& {Ichiki}}]{Shimabukuro_2016}
{Shimabukuro}, H., {Yoshiura}, S., {Takahashi}, K., {Yokoyama}, S., \& {Ichiki}, K. 2016, \mnras, 458, 3003, \dodoi{10.1093/mnras/stw482}

\bibitem[{Shimabukuro {et~al.}(2017)Shimabukuro, Yoshiura, Takahashi, Yokoyama, \& Ichiki}]{shiam2017}
Shimabukuro, H., Yoshiura, S., Takahashi, K., Yokoyama, S., \& Ichiki, K. 2017, Mon. Not. Roy. Astron. Soc., 468, 1542, \dodoi{10.1093/mnras/stx530}

\bibitem[{{Swarup} {et~al.}(1991){Swarup}, {Ananthakrishnan}, {Kapahi}, {Rao}, {Subrahmanya}, \& {Kulkarni}}]{Swarup_1991}
{Swarup}, G., {Ananthakrishnan}, S., {Kapahi}, V.~K., {et~al.} 1991, Current Science, 60, 95

\bibitem[{{Thompson} {et~al.}(2017){Thompson}, {Moran}, \& {Swenson}}]{Thompson_1986}
{Thompson}, A.~R., {Moran}, J.~M., \& {Swenson}, G.~W. 2017, {Interferometry and synthesis in radio astronomy} (Springer Cham)

\bibitem[{{Tingay} {et~al.}(2013){Tingay}, {Goeke}, {Bowman}, {Emrich}, {Ord}, {Mitchell}, {Morales}, {Booler}, {Crosse}, {Wayth}, {Lonsdale}, {Tremblay}, {Pallot}, {Colegate}, {Wicenec}, {Kudryavtseva}, {Arcus}, {Barnes}, {Bernardi}, {Briggs}, {Burns}, {Bunton}, {Cappallo}, {Corey}, {Deshpande}, {Desouza}, {Gaensler}, {Greenhill}, {Hall}, {Hazelton}, {Herne}, {Hewitt}, {Johnston-Hollitt}, {Kaplan}, {Kasper}, {Kincaid}, {Koenig}, {Kratzenberg}, {Lynch}, {Mckinley}, {Mcwhirter}, {Morgan}, {Oberoi}, {Pathikulangara}, {Prabu}, {Remillard}, {Rogers}, {Roshi}, {Salah}, {Sault}, {Udaya-Shankar}, {Schlagenhaufer}, {Srivani}, {Stevens}, {Subrahmanyan}, {Waterson}, {Webster}, {Whitney}, {Williams}, {Williams}, \& {Wyithe}}]{tingay13}
{Tingay}, S.~J., {Goeke}, R., {Bowman}, J.~D., {et~al.} 2013, \pasa, 30, e007, \dodoi{10.1017/pasa.2012.007}

\bibitem[{{Tiwari} {et~al.}(2022){Tiwari}, {Shaw}, {Majumdar}, {Kamran}, \& {Choudhury}}]{tiwari_2022}
{Tiwari}, H., {Shaw}, A.~K., {Majumdar}, S., {Kamran}, M., \& {Choudhury}, M. 2022, JCAP, 2022, 045, \dodoi{10.1088/1475-7516/2022/04/045}

\bibitem[{Trott {et~al.}(2019)}]{trott2019}
Trott, C.~M., {et~al.} 2019, Publ. Astron. Soc. Austral., 36, e023, \dodoi{10.1017/pasa.2019.15}

\bibitem[{{van Haarlem, M. P.} {et~al.}(2013){van Haarlem, M. P.}, {Wise, M. W.}, {Gunst, A. W.}, {Heald, G.}, {McKean, J. P.}, {Hessels, J. W. T.}, {de Bruyn, A. G.}, {Nijboer, R.}, {Swinbank, J.}, {Fallows, R.}, {Brentjens, M.}, {Nelles, A.}, {Beck, R.}, {Falcke, H.}, {Fender, R.}, {H\"orandel, J.}, {Koopmans, L. V. E.}, {Mann, G.}, {Miley, G.}, {R\"ottgering, H.}, {Stappers, B. W.}, {Wijers, R. A. M. J.}, {Zaroubi, S.}, {van den Akker, M.}, {Alexov, A.}, {Anderson, J.}, {Anderson, K.}, {van Ardenne, A.}, {Arts, M.}, {Asgekar, A.}, {Avruch, I. M.}, {Batejat, F.}, {B\"ahren, L.}, {Bell, M. E.}, {Bell, M. R.}, {van Bemmel, I.}, {Bennema, P.}, {Bentum, M. J.}, {Bernardi, G.}, {Best, P.}, {B\^{\i}rzan, L.}, {Bonafede, A.}, {Boonstra, A.-J.}, {Braun, R.}, {Bregman, J.}, {Breitling, F.}, {van de Brink, R. H.}, {Broderick, J.}, {Broekema, P. C.}, {Brouw, W. N.}, {Br\"uggen, M.}, {Butcher, H. R.}, {van Cappellen, W.}, {Ciardi, B.}, {Coenen, T.}, {Conway, J.}, {Coolen, A.}, {Corstanje, A.}, {Damstra, S.}, {Davies,
  O.}, {Deller, A. T.}, {Dettmar, R.-J.}, {van Diepen, G.}, {Dijkstra, K.}, {Donker, P.}, {Doorduin, A.}, {Dromer, J.}, {Drost, M.}, {van Duin, A.}, {Eisl\"offel, J.}, {van Enst, J.}, {Ferrari, C.}, {Frieswijk, W.}, {Gankema, H.}, {Garrett, M. A.}, {de Gasperin, F.}, {Gerbers, M.}, {de Geus, E.}, {Grie\ss{}meier, J.-M.}, {Grit, T.}, {Gruppen, P.}, {Hamaker, J. P.}, {Hassall, T.}, {Hoeft, M.}, {Holties, H. A.}, {Horneffer, A.}, {van der Horst, A.}, {van Houwelingen, A.}, {Huijgen, A.}, {Iacobelli, M.}, {Intema, H.}, {Jackson, N.}, {Jelic, V.}, {de Jong, A.}, {Juette, E.}, {Kant, D.}, {Karastergiou, A.}, {Koers, A.}, {Kollen, H.}, {Kondratiev, V. I.}, {Kooistra, E.}, {Koopman, Y.}, {Koster, A.}, {Kuniyoshi, M.}, {Kramer, M.}, {Kuper, G.}, {Lambropoulos, P.}, {Law, C.}, {van Leeuwen, J.}, {Lemaitre, J.}, {Loose, M.}, {Maat, P.}, {Macario, G.}, {Markoff, S.}, {Masters, J.}, {McFadden, R. A.}, {McKay-Bukowski, D.}, {Meijering, H.}, {Meulman, H.}, {Mevius, M.}, {Middelberg, E.}, {Millenaar, R.}, {Miller-Jones, J.
  C. A.}, {Mohan, R. N.}, {Mol, J. D.}, {Morawietz, J.}, {Morganti, R.}, {Mulcahy, D. D.}, {Mulder, E.}, {Munk, H.}, {Nieuwenhuis, L.}, {van Nieuwpoort, R.}, {Noordam, J. E.}, {Norden, M.}, {Noutsos, A.}, {Offringa, A. R.}, {Olofsson, H.}, {Omar, A.}, {Orr\'u, E.}, {Overeem, R.}, {Paas, H.}, {Pandey-Pommier, M.}, {Pandey, V. N.}, {Pizzo, R.}, {Polatidis, A.}, {Rafferty, D.}, {Rawlings, S.}, {Reich, W.}, {de Reijer, J.-P.}, {Reitsma, J.}, {Renting, G. A.}, {Riemers, P.}, {Rol, E.}, {Romein, J. W.}, {Roosjen, J.}, {Ruiter, M.}, {Scaife, A.}, {van der Schaaf, K.}, {Scheers, B.}, {Schellart, P.}, {Schoenmakers, A.}, {Schoonderbeek, G.}, {Serylak, M.}, {Shulevski, A.}, {Sluman, J.}, {Smirnov, O.}, {Sobey, C.}, {Spreeuw, H.}, {Steinmetz, M.}, {Sterks, C. G. M.}, {Stiepel, H.-J.}, {Stuurwold, K.}, {Tagger, M.}, {Tang, Y.}, {Tasse, C.}, {Thomas, I.}, {Thoudam, S.}, {Toribio, M. C.}, {van der Tol, B.}, {Usov, O.}, {van Veelen, M.}, {van der Veen, A.-J.}, {ter Veen, S.}, {Verbiest, J. P. W.}, {Vermeulen, R.}, {Vermaas,
  N.}, {Vocks, C.}, {Vogt, C.}, {de Vos, M.}, {van der Wal, E.}, {van Weeren, R.}, {Weggemans, H.}, {Weltevrede, P.}, {White, S.}, {Wijnholds, S. J.}, {Wilhelmsson, T.}, {Wucknitz, O.}, {Yatawatta, S.}, {Zarka, P.}, {Zensus, A.}, \& {van Zwieten, J.}}]{haarlem}
{van Haarlem, M. P.}, {Wise, M. W.}, {Gunst, A. W.}, {et~al.} 2013, A\&A, 556, A2, \dodoi{10.1051/0004-6361/201220873}

\bibitem[{{Watkinson} {et~al.}(2022){Watkinson}, {Greig}, \& {Mesinger}}]{Watkinson_2022}
{Watkinson}, C.~A., {Greig}, B., \& {Mesinger}, A. 2022, \mnras, 510, 3838, \dodoi{10.1093/mnras/stab3706}

\bibitem[{{Watkinson} {et~al.}(2021){Watkinson}, {Trott}, \& {Hothi}}]{Watkinson_2021}
{Watkinson}, C.~A., {Trott}, C.~M., \& {Hothi}, I. 2021, \mnras, 501, 367, \dodoi{10.1093/mnras/staa3677}

\bibitem[{{Yoshiura} {et~al.}(2015){Yoshiura}, {Shimabukuro}, {Takahashi}, {Momose}, {Nakanishi}, \& {Imai}}]{Yoshiura_2015}
{Yoshiura}, S., {Shimabukuro}, H., {Takahashi}, K., {et~al.} 2015, \mnras, 451, 266, \dodoi{10.1093/mnras/stv855}

\end{thebibliography}
\bibliographystyle{aasjournal}

\label{lastpage}
\end{document}